\documentclass[12pt,fleqn]{article}
\usepackage{cite}
\usepackage{amssymb,amsmath,epsfig}
\newcommand{\newsection}[1]{\section{#1}\setcounter{equation}{0}}
\renewcommand{\thefootnote}{\fnsymbol{footnote}}

\def \be{\begin{equation}}
\def \ee{\end{equation}}
\def \bea{\begin{eqnarray}}
\def \eea{\end{eqnarray}}
\def \ben{\begin{enumerate}}
\def \een{\end{enumerate}}
\def \bit{\begin{itemize}}
\def \eit{\end{itemize}}
\def \Bbar{\overline{\kern -0.24em B}}
\def \all{A_{0L}}
\def \alr{A_{0R}}
\def \apl{A_{\| L}}
\def \apr{A_{\| R}}
\def \appl{A_{\bot L}}
\def \appr{A_{\bot R}}
\def \al{A_0}
\def \ap{A_{\|}}
\def \app{{A}_{\bot}}
\def \mhat{\hat{m}_\kstar}
\def \eff{\mathrm{eff}}

\def \kstar{{K^*}}
\def \trans{{T}}

\def \sh{\hat{s}}
\def \Im{{\text{Im}}}
\def \Re{{\text{Re}}}
\def \GeV{{\text{GeV}}}

\def \bm{\boldmath}

\def \braket#1#2#3{\langle #1|#2| #3\rangle}

\def \ea{{\it et al.}}
\def \eq#1{Eq.~(\ref{#1})}
\def \eqs#1#2{Eqs.~(\ref{#1})--(\ref{#2})}
\def \fig#1{Fig.~\ref{#1}}

\def \nnu{\nonumber}
\def \O{{\cal O}}
\def \ol#1{\overline{#1}}

\def \rf{Ref.~\cite}
\def \rfs{Refs.~\cite}
\def \sec#1{Sec.~\ref{#1}}

\def \cseff{C_7^{\text{eff}}}
\def \cseffP{{C_7^{\text{eff}}}^\prime}
\def \ceff{C_9^{\text{eff}}}

\def \cten{C_{10}}

\def \a{\alpha}

\def \g{\gamma}
\def \G{\Gamma}

\def \m{\mu}
\def\euro#1#2#3{{Eur. Phys. J. C} {\bf #1}, #3 (#2)}
\def\eurodirect#1#2#3{{EPJdirect} {\bf C#1}, #3 (#2)}

\def\ibid#1#2#3{{\it ibid.\/}~{\bf#1}, #3 (#2)}
\def\ib#1#2#3{{\bf#1}, #3 (#2)}
\def\jhep#1#2#3{{J.~High Energy Phys.} {\bf #1}, #3 (#2)}

\def\np#1#2#3{{Nucl.~Phys.}~{\bf B#1}, #3 (#2)}

\def\pl#1#2#3{{Phys.~Lett. B}~{\bf #1}, #3 (#2)}

\def\prd#1#2#3{{Phys.~Rev. D}~{\bf #1}, #3 (#2)}
\def\prl#1#2#3{{Phys.~Rev.~Lett.}~{\bf #1}, #3 (#2)}

\def\rmp#1#2#3{{Rev. Mod. Phys.} {\bf #1}, #3 (#2)}
\def\zpc#1#2#3{{Z.~Phys. C}~{\bf #1}, #3 (#2)}
\begin{document}
\begin{titlepage}
\begin{flushright}
UAB-FT 560\\
hep-ph/0502060\\
February 2005
\end{flushright}
\vskip 0.2in
\begin{center}
\setlength{\baselineskip}{0.35in} 
%
%

{\bm\bf\Large Probing New Physics Via the Transverse Amplitudes of 
\bm${B}^0\to K^{*0}(\to K^- \pi^+) l^+l^-$ at Large Recoil}  \\[1.2cm]
\setlength {\baselineskip}{0.25in}
{\textsc{Frank Kr\"uger$^{1, }$\footnote{E-mail address: frank.krueger@fhm.edu} and
 Joaquim Matias$^{2, }$\footnote{E-mail address:  matias@ifae.es}}}\\[1mm]
\emph{$^1$Munich University of Applied Sciences, D-80335 M\"unchen, Germany}\\
\emph{$^2$IFAE, Universitat Aut\`onoma de Barcelona, 08193
Bellaterra, Barcelona, Spain}
\end{center}
\begin{abstract}
We perform an analysis of the  $K^*$ polarization states in  
the exclusive $B$ meson decay ${B}^0\to~K^{*0}(\to K^- \pi ^+)l^+l^-$ 
($l=e, \mu, 
\tau$) 
in the low dilepton mass region, where the final vector meson has a large energy. 
Working in the transversity basis, we study
various observables that involve the $K^*$ 
spin amplitudes $\app$, $\ap$, $\al$ by exploiting the 
heavy-to-light form factor relations in the heavy quark and large-$E_{K^*}$ limit. 
We find that at leading order in $1/m_b $ and $\alpha_s$ 
the form-factor dependence of the asymmetries that 
involve transversely polarized $K^*$ completely drops out.~At next-to-leading logarithmic order,
including factorizable and non-factorizable 
corrections, the theoretical errors for the transverse asymmetries turn out to be small in the standard model (SM). 
Integrating over the lower part of the dimuon mass region,
and varying the theoretical input parameters, the SM predicts
${\mathcal A}_T^{(1)}=0.9986\pm 0.0002$ and  
${\mathcal A}_T^{(2)}=-0.043\pm 0.003$. In addition, the 
longitudinal and transverse polarization 
fractions are found to be $(69\pm 3)\%$ and $(31\pm 3)\%$ respectively, so that
$\Gamma_L/\Gamma_T = 2.23\pm 0.31$.
Beyond the SM, we focus on new physics that mainly gives sizable 
contributions to the coefficients $C_7^{\mathrm{eff}(\prime)}$ of the 
electromagnetic dipole operators. Taking 
into account experimental data on rare $B$ decays, we find large effects of new 
physics in the transverse asymmetries. Furthermore, we show that a measurement of 
longitudinal and transverse polarization fractions
will provide complementary information on physics beyond the SM.

\noindent
PACS number(s): 13.20.He, 13.25.Hw
\end{abstract}
\end{titlepage}
%
%
\renewcommand{\thefootnote}{\arabic{footnote}}
\setcounter{footnote}{0}
\newsection{Introduction}
The decay ${B^0}\to K^{*0}(\to K^- \pi ^+)l^+l^-$, where
$l$ stands for $e,\mu,\tau$, is interesting as a 
testing ground for the standard model (SM) and
its extensions \cite{dmitri,FK:etal,CS:etal,chen:geng:Tviolation,
BtoKpill:gamma,ali:safir,thq,ali:etal,Ali:2002jg}. 
In particular, it has been shown that a study of 
the angular distribution of the four-body final
state allows to search for right-handed currents \cite{dmitri,CS:etal} and
provides additional information on CP violation \cite{FK:etal}.

Since the exclusive $B^0\to K^{*0} l^+l^-$ decay involves 
the heavy-to-light transition form factors 
parametrizing the hadronic matrix elements, it usually suffers from large theoretical 
uncertainties, which amount to $\sim 30\%$ on the branching ratio 
\cite{ali:etal,Ali:2002jg}.~However, the theoretical errors 
can be reduced by exploiting   relations between the form factors 
that emerge in the limit where the initial hadron is 
heavy and the final meson has a large energy \cite{large:energy:limit}. In this case, 
the seven a priori independent $B\to K^*$ transition form factors   
can be expressed through merely two universal form factors  at leading power in  
$1/m_b$ and $\alpha_s$ \cite{large:energy:limit}. 
While this reduces the hadronic uncertainties in the 
calculation of exclusive $B$ decays, it restricts the validity of the theoretical 
predictions to the dilepton  mass region below the $J/\psi$ mass. 
Corrections to the heavy quark and large energy limit, at next-to-leading logarithmic (NLL) order, have been computed in 
\cite{beneke:feldmann,beneke:etal}
including factorizable and non-factorizable contributions. 
(The corrections to the heavy-to-light form factors at large recoil energy can 
be calculated systematically within the soft-collinear effective theory \cite{SCET}.)

In this paper, we study various observables that involve different combinations of 
$K^*$ spin amplitudes, whose moduli and phases can be extracted from an analysis of the 
angular distribution of ${B}^0\to K^{*0}(\to K^- \pi ^+)l^+l^-$ \cite{FK:etal}.
In order to reduce the uncertainties due to the hadronic form factors, we concentrate
on quantities that contain only ratios of $K^*$ polarization amplitudes. 
Special emphasis is put on those observables that involve transversely 
polarized $K^*$, which turn out to be largely independent of 
the hadronic form factors, even after including NLL corrections.
We also make use of the NLL order corrections \cite{beneke:feldmann,beneke:etal},
as a first approach, to study the robustness of the set of observables 
analysed below. Subleading effects in the heavy quark expansion 
\cite{thq,kn} will be 
included elsewhere.

Our paper is organized as follows. In \sec{theo:setup} we setup our 
theoretical framework. Section \ref{ang:dist} contains  the angular distribution in 
the transversity basis and the polarization amplitudes of the final vector meson. 
In \sec{sec:Kpol} we discuss  various observables
that are sensitive to the polarization states of $K^*$. 
We study in detail the 
implications of factorizable and non-factorizable corrections to these 
observables in the SM.
The impact of new physics is investigated in a 
model-independent manner. In particular, the implications 
of right-handed currents in the low dilepton invariant mass region are examined. 
Our summary and conclusions can be found in \sec{summary}. 
For the paper to be self-contained,  
we provide in the appendix
the angular distribution of ${B}^0\to K^{*0}(\to K^- \pi ^+)l^+l^-$ 
including lepton-mass effects.

\newsection{Theoretical framework}\label{theo:setup}
We begin with the matrix element of the decay 
${B^0}\to K^{*0}(\to K^- \pi ^+)l^+l^-$, which may be obtained by using the
effective  Hamiltonian describing the $b\to s l^+l^-$ transition 
\cite{dmitri,FK:etal,CS:etal}. It can be written as \cite{wilson:coeffs:SM}
\be\label{heff}
{\mathcal{H}}_{\eff}=-\frac{4 G_F}{\sqrt{2}} V_{tb}^{} V_{ts}^* 
\sum_{i=1}^{10}
[C_i (\mu) {\mathcal{O}}_{i}(\mu)  + C_i^\prime (\mu) {\mathcal{O}}_{i}^\prime (\mu)],
\ee
where $C_i^{(\prime)}(\mu)$ and ${\mathcal{O}}_{i}^{(\prime)}(\mu)$ are
the Wilson coefficients and local operators respectively. For a complete set of 
operators in the SM (i.e.~${\mathcal{O}}_{i}$'s) and beyond, we refer 
to \cite{wilson:coeffs:SM,LRmodel,Borzumati:1999qt}.

In our subsequent analysis, we concentrate on
\bea\label{operator:basis}
{\mathcal{O}}_{7} = \frac{e}{16\pi^2} m_b
(\bar{s} \sigma_{\mu \nu} P_R b) F^{\mu \nu}, \quad
{\mathcal{O}}_{9} = \frac{e^2}{16\pi^2} 
(\bar{s} \gamma_{\mu} P_L b)(\bar{l} \gamma^\mu l), \quad
{\mathcal{O}}_{10}=\frac{e^2}{16\pi^2}
(\bar{s}  \gamma_{\mu} P_L b)(  \bar{l} \gamma^\mu \gamma_5 l),\nnu\\
\eea
where $P_{L,R}= (1\mp \g_5)/2$ and $m_b \equiv   m_b(\mu)$ is the running mass in the 
$\ol{\mbox{MS}}$ scheme. As for the primed operators \cite{LRmodel,Borzumati:1999qt}, 
we restrict ourselves to   
\bea
{\mathcal{O}}_{7}^\prime = \frac{e}{16\pi^2} m_b
(\bar{s} \sigma_{\mu \nu} P_L b) F^{\mu \nu}.
\eea
(Below we show  how to include the chiral partners of 
${\cal O}_{9,10}$ in our analysis.)
Since there are no right-handed currents in the SM,\footnote{Throughout
this paper we neglect the strange quark mass and do not consider 
CP violation.}
the coefficient accompanying ${\mathcal{O}}_{7}^\prime$ is non-zero 
only in certain extensions of the 
SM such as the left-right model \cite{LRmodel} and the unconstrained 
supersymmetric standard model \cite{Borzumati:1999qt}.

\subsection{Matrix element}
Given the above Hamiltonian, the matrix element can be written as 
\bea\label{matrix:ele}
{\mathcal M} &=& \frac{G_F\a}{\sqrt{2}\pi}V_{tb}^{}V_{ts}^*\bigg\{
\bigg[\ceff\braket{K \pi}{(\bar{s}\g^{\mu}P_Lb)}{B} -\frac{2m_b}{q^2}
\braket{K\pi}{\bar{s}i\sigma^{\mu\nu}q_{\nu}(\cseff P_R+\cseffP  P_L)b}{B}\bigg]
(\bar{l}\g_{\m}l)\nnu\\
&+& \cten\braket{K\pi}{(\bar{s}\g^{\mu}P_Lb)}{B}(\bar{l}\g_{\mu}\g_5 l)\bigg \},
\eea
where $q$ is the four-momentum of the lepton pair. Explicit expressions for 
the short-distance coefficients, including
next-to-next-to-leading logarithmic (NNLL) corrections, can be found in
 \rfs{NNLO:OB,NNLO:ME:formulae,Gambino:2003zm}. 

The hadronic part of the matrix element describing the $B\to K\pi$ transition 
can be para\-me\-trized in terms of $B\to K^*$  
form factors by means of a narrow-width approximation \cite{FK:etal}.
The relevant form factors are defined as  \cite{wirbel:etal,ali:etal}:
\bea\label{ff:btokstar:vector}
\lefteqn{\braket{\kstar(p_{\kstar})}{\bar{s}\g_{\m}P_{L,R} b}{{B}(p)} 
= i\epsilon_{\mu\nu\alpha\beta} \epsilon^{\nu*}p^{\alpha}q^{\beta} \frac{V(s)}{m_B+m_{\kstar}}
\mp\frac{1}{2}\bigg\{\epsilon_{\m}^*(m_B+ m_{\kstar})A_1(s)}\nnu\\
&-&(\epsilon^*\cdot q)(2p -q)_\mu \frac{A_2(s)}{m_B+m_{\kstar}} -
\frac{2m_{\kstar}}{s} (\epsilon^*\cdot q) [A_3(s)- A_0(s)]q_\mu\bigg\},
\eea
where 
\bea
A_3 (s) = \frac{m_B+m_{\kstar}}{2m_{\kstar}} A_1(s) - \frac{m_B-m_{\kstar}}{2m_{\kstar}}A_2(s),
\eea
and 
\bea\label{ff:btokstar:tensor}
\lefteqn{\braket{\kstar(p_{\kstar})}{\bar{s}i \sigma_{\mu\nu}q^{\nu}P_{R,L} b}{{B}(p)}
= - i\epsilon_{\mu\nu\alpha\beta} \epsilon^{\nu*}p^{\alpha}q^{\beta}T_1(s)
\pm\frac{1}{2}\bigg\{[\epsilon_{\mu}^*(m_B^2-m_{\kstar}^2)}\nnu\\
&-& (\epsilon^*\cdot q)(2p-q)_{\mu}] T_2(s)
+ (\epsilon^*\cdot q)
\bigg[ q_\mu - \frac{s}{m_B^2-m_{\kstar}^2} (2p - q)_\mu\bigg]T_3(s)\bigg\}.
\eea
In the above,  $q = p_{l^+}+ p_{l^-}$, $s = q^2$, and $\epsilon^{\m}$ is the $K^*$ polarization vector.

\subsection{Heavy-to-light form factors at large recoil}
As we have already mentioned, interesting relations between the 
hadronic form factors emerge in the limit
where the initial hadron is heavy and the final meson has a large
energy \cite{large:energy:limit}. In this case, the form factors can be 
expanded in the small ratios $\Lambda_{\mathrm{QCD}}/m_b$ and
$\Lambda_{\mathrm{QCD}}/E$, where $E$ is the energy of
the light meson.  Neglecting  corrections of order $1/m_b$ and $\alpha_s$, 
the seven a priori independent $B\to K^*$ form factors  
in Eqs.~(\ref{ff:btokstar:vector}) and (\ref{ff:btokstar:tensor}) 
reduce to two universal form factors $\xi_{\bot}$ and  $\xi_{\|}$
\cite{large:energy:limit,beneke:feldmann}:\footnote{Following \rf{beneke:feldmann},
the longitudinal form factor $\xi_\|$ is related to that of 
\rf{large:energy:limit} by $\xi_{\|} = (m_\kstar/E_\kstar) \zeta_{\|}$.} 
\begin{subequations}\label{form:factor:relations:LEL}
\be
A_1(s)= \frac{2 E_\kstar}{m_B + m_\kstar}\xi_{\bot}(E_\kstar),
\ee
\be
A_2(s)= \frac{m_B}{m_B-m_\kstar}
\bigg[\xi_{\bot}(E_\kstar)- \xi_{\|}(E_\kstar)\bigg],
\ee
\be
A_0(s) = \frac{E_\kstar}{m_\kstar} \xi_{\|}(E_\kstar),
\ee
\be
V(s)= \frac{m_B+m_\kstar}{m_B}\xi_{\bot}(E_\kstar),
\ee
\be
T_1(s)=\xi_{\bot}(E_\kstar),
\ee
\be
T_2(s)=\frac{2 E_\kstar}{m_B}\xi_{\bot}(E_\kstar),
\ee
\be
T_3(s)=\xi_{\bot}(E_\kstar) - \xi_{\|}(E_\kstar).
\ee
\end{subequations}%
Here, $E_\kstar$ is the energy of the final vector meson in the ${B}$ 
rest frame,  
\be
E_\kstar  \simeq  \frac{m_B}{2}\left(1-\frac{s}{m_B^2}\right).
\ee
Since the theoretical predictions are restricted to the kinematic region 
in which the energy of the $K^*$ is of the order of the heavy quark mass 
(i.e.~$s \ll m_B^2$), we confine our analysis 
to the dilepton mass in the range  
$2m_l\leqslant M_{l^+l^-}\leqslant 2.5\ \GeV$.  
The relations in Eqs.~(\ref{form:factor:relations:LEL}) are valid for the 
soft contribution to the form factors at large recoil, and are violated by 
symmetry breaking corrections of order 
$\alpha_s$ and $1/m_b$. 
The corrections at first order in 
$\alpha_s$
have been computed in  
\cite{beneke:feldmann,beneke:etal} and will be discussed in 
more detail in \sec{SM:prediction}.
However, it should be noted that the ratios $A_1/V$ and 
$T_1/T_2$ do not receive 
$\alpha_s$ corrections to leading power in $1/E_{K^*}$ 
\cite{beneke:feldmann,Burdman:2000ku}. This in turn leads to a specific behaviour 
of the amplitudes describing the transverse polarization states of the $K^*$
in the heavy quark and large energy limit, as explained below.

\newsection{Angular distribution  and transversity amplitudes}\label{ang:dist}
Assuming the $K^*$ to be on the mass shell, the decay
 ${B^0}\to K^{*0}(\to K^- \pi ^+)l^+l^-$ is  completely 
described by four independent kinematic variables; namely, the lepton-pair invariant mass, $s$, 
and the three angles $\theta_l$, $\theta_{K^*}$, $\phi$.  
In terms of these variables, the differential decay rate can be written as 
\cite{FK:etal}
\bea\label{diff:decay:rate}
\frac{d^4\G}{ds\,d\cos\theta_l\,  d\cos\theta_{K^*}\, d\phi}
= \frac{9}{32 \pi}\sum_{i=1}^9 I_i(s,\theta_{K^*})f_i(\theta_l,\phi),
\eea
where $I_i$ depend on products of the four $K^*$ spin amplitudes   
$\app$, $\ap$, $\al$, $A_t$, and $f_i$ are 
the corresponding angular distribution functions (see Appendix \ref{kin} for details).
Note that $A_t$ is related  to the time-like component of the 
virtual $K^*$, which does not contribute in the case of massless leptons.

Given the matrix element in \eq{matrix:ele}, we obtain for the transversity amplitudes
\be\label{a_perp}
A_{\bot L,R}=N \sqrt{2} \lambda^{1/2}\bigg[
(\ceff\mp\cten)\frac{V(s)}{m_B +m_\kstar}+\frac{2m_b}{s} (\cseff + \cseffP) 
T_1(s)\bigg], 
\ee
\be\label{a_par}
A_{\| L,R}= - N \sqrt{2}(m_B^2- m_\kstar^2)\bigg[(\ceff\mp \cten)
\frac{A_1 (s)}{m_B-m_\kstar} 
+\frac{2 m_b}{s} (\cseff - \cseffP) T_2(s)\bigg],
\ee
\bea\label{a_long}
A_{0L,R}&=&-\frac{N}{2m_\kstar\sqrt{s}}\bigg[
(\ceff\mp \cten)\bigg\{(m_B^2-m_\kstar^2 -s)(m_B+m_\kstar)A_1(s)
-\lambda \frac{A_2(s)}{m_B +m_\kstar}\bigg\}\nnu\\
&+&{2m_b}(\cseff - \cseffP) \bigg\{
 (m_B^2+3m_\kstar^2 -s)T_2(s)
-\frac{\lambda}{m_B^2-m_\kstar^2} T_3(s)\bigg\}\bigg],
\eea
\bea\label{a_t}
A_t= \frac{2 N }{\sqrt{s}}\lambda^{1/2} C_{10}A_0(s) ,
\eea
which are related to the helicity amplitudes used, e.g., in 
\cite{dmitri,CS:etal,ali:safir} through 
\be\label{hel:trans} 
A_{\bot,\|} = (H_{+1}\mp H_{-1})/\sqrt{2}, \quad A_0=H_0, \quad A_t=H_t.
\ee 
In the above formulae, 
$\lambda= m_B^4  + m_{K^*}^4 + s^2 - 2 (m_B^2 m_{K^*}^2+ m_{K^*}^2 s  + m_B^2 s)$ and 
\be
N=\left[\frac{G_F^2 \a^2}{3\cdot 2^{10}\pi^5 m_B^3}
|V_{tb}^{}V_{ts}^{\ast}|^2 s \lambda^{1/2}
\left(1-\frac{4 m_l^2}{s}\right)^{1/2}\right]^{1/2}.
\ee
Note that the contributions of  the chirality-flipped operators 
${\mathcal{O}}_{9,10}^\prime ={\mathcal{O}}_{9,10} (P_L\to P_R)$
can be included in the above amplitudes by the replacements
$C^{(\mathrm{eff})}_{9,10}\to C^{(\mathrm{eff})}_{9,10}+ 
C^{(\mathrm{eff})\prime}_{9,10}$ in \eq{a_perp}, 
$C_{9,10}^{(\mathrm{eff})}\to C^{(\mathrm{eff})}_{9,10}- C^{(\mathrm{eff})\prime}_{9,10}$ in Eqs.~(\ref{a_par}) and (\ref{a_long}),
and $C_{10}\to C_{10}- C_{10}^\prime$ in \eq{a_t}.

\subsection{Transversity amplitudes at large recoil}
The transversity amplitudes in \eqs{a_perp}{a_t} 
take a particularly simple form in the heavy quark and large energy limit.~In fact, exploiting the form factor relations in Eqs.~(\ref{form:factor:relations:LEL}), 
we obtain at leading order in $1/m_b$ and $\alpha_s$ 
\be\label{LEL:tranversity:perp}
A_{\bot L,R}=  \sqrt{2} N m_B(1- \sh)\bigg[
(\ceff\mp\cten)+\frac{2\hat{m}_b}{\sh} (\cseff + \cseffP) 
\bigg]\xi_{\bot}(E_\kstar),
\ee 
\be\label{LEL:tranversity:par}
A_{\| L,R}= -\sqrt{2} N m_B (1-\sh)\bigg[(\ceff\mp \cten) 
+\frac{2\hat{m}_b}{\sh}(\cseff -\cseffP) \bigg]\xi_{\bot}(E_\kstar),
\ee 
\bea\label{LEL:tranversity:zero}
A_{0L,R}= -\frac{Nm_B }{2 \hat{m}_\kstar \sqrt{\sh}} (1-\sh)^2\bigg[(\ceff\mp \cten) + 2
\hat{m}_b (\cseff -\cseffP) \bigg]\xi_{\|}(E_\kstar),
\eea
\bea\label{LEL:tranversity:t}
A_t= \frac{N m_B }{\hat{m}_\kstar \sqrt{\sh}}(1- \sh)^2 C_{10}\xi_{\|}(E_\kstar),
\eea
with $\sh =  s/m_B^2$, $\hat{m}_i =  m_i/m_B$. In writing 
Eqs.~(\ref{LEL:tranversity:perp})--(\ref{LEL:tranversity:t}) we have 
dropped terms of $O(\hat{m}_{K^*}^2)$. From inspection
of these formulae, we infer the following features.

(i)  Within the SM, we recover the naive quark-model prediction of 
$A_{\bot}= - A_{\|}$ \cite{quark:model,soares}  in the 
$m_B\to \infty$ and $E_\kstar \to \infty$ limit (equivalently $\mhat^2\to 0$). 
In this case, the $s$ quark is produced in helicity 
$-{1}/{2}$ by weak interactions 
in the limit $m_s\to 0$, which is not affected by strong 
interactions in the massless case \cite{Burdman:2000ku}. Thus, 
the strange quark combines with a light quark to form a $K^*$ with helicity  
either $-1$ or $0$ but not $+1$. Consequently,  the SM predicts at quark level
$H_{+1}=0$, and hence $A_{\bot}= - A_{\|}$ [cf.~\eq{hel:trans}], which is revealed as 
$|H_{-1}|\gg |H_{+1}|$ (or $A_{\bot}\approx  - A_{\|}$) at the hadron level.

(ii) The longitudinal and time-like (transverse) polarizations of the 
$K^*$ involve only the universal
form factors  $\xi_{\|}$ ($\xi_{\bot}$) in the $m_B\to \infty$ and $E_\kstar \to
\infty$ limit. 

\newsection{\bm$K^*$ polarization as a probe of new physics}\label{sec:Kpol}
\subsection{Observables}
The study of the angular distribution in the decay 
${B^0}\to K^{*0}(\to K^- \pi ^+)l^+l^-$ allows a determination of the $K^*$
spin amplitudes 
along with  their relative phases
(see Appendix \ref{kin}). In order to minimize the theoretical uncertainties due to the 
hadronic form factors, we consider only those observables that involve 
ratios of amplitudes.  (For a discussion of the error on the 
decay amplitudes of $B^0\to K^{*0} l^+l^-$, we refer to \rf{ali:safir}.) 

Introducing the shorthand notation
\bea
A_i A^*_j\equiv A^{}_{i L}(s) A^*_{jL}(s)+ A^{}_{iR}(s) A^*_{jR}(s) \quad (i,j  = 0, \|, \perp),
\eea
we investigate the following observables.

(i) Transverse asymmetries 
\be\label{def:asymmetries}
A^{(1)}_{\trans}({s})=\frac{-2\Re(\ap^{}\app^*)}{|\app|^2 + |\ap|^2},\quad
A^{(2)}_{\trans}({s})=\frac{|\app|^2 - |\ap|^2}{|\app|^2 + |\ap|^2}.
\ee

(ii) $K^*$ polarization parameter
\bea\label{def:pol:param}
\alpha_{K^*}(s) = \frac{2|{ A}_0|^2}{|{A}_{\|}|^2 + |{A}_{\perp}|^2}-1.
\eea
For final states with $l=e$ or $\mu$ it can be directly 
determined from the two-dimensional differential decay rate
$d\Gamma/(ds\, d\cos\theta_{K^*}) \propto [1 + \alpha_{K^*}(s) \cos^2 \theta_{K^*}]$, 
since the corresponding lepton-mass corrections are negligibly small.

(iii) Fraction of $K^*$ polarization 
\bea\label{def:frac:pol}
F_L(s)
= \frac{|{A}_0|^2}{|{ A}_0|^2 + |{A}_{\|}|^2
+ |A_\perp|^2}, 
\quad F_T(s) = \frac{|{A}_\bot|^2+|{A}_\||^2}{|{ A}_0|^2 + |{A}_{\|}|^2 + |A_\perp|^2},
\eea
so that $\alpha_{K^*}=2F_L/F_T-1$.

(iv) Integrated quantities ${\mathcal A}^{(1)}_{\trans}$, ${\mathcal A}^{(2)}_{\trans}$, $\mbox{\bm $\alpha$}_{K^*}$, and  ${\mathcal F}_{L,T}$, which are obtained from the ones 
 above by integrating numerator and denominator separately over the dilepton invariant 
mass. 

\subsection{SM prediction for the \bm $K^*$ polarization}\label{SM:prediction}
In this section we perform a detailed analysis of the previously defined observables 
within the SM. 
Following the work of Beneke \emph{et al.}
\cite{beneke:etal}, we include factorizable and 
non-factorizable corrections at NLL order. Since 
these results are  applicable only to the region where $s \lesssim 4m_c^2$, we 
consider in the remainder of this paper muons in the final state with 
$2m_\mu \leqslant M_{\mu^+\mu^-} \leqslant 2.5\ \GeV $. 

To include the NLL corrections to the transversity amplitudes 
 in the SM, we  set $\cseffP=0$  in \eqs{a_perp}{a_long} and replace
\be\label{subs:Tis}
\cseff T_i \rightarrow {\cal T}_i, \quad \ceff \rightarrow C_9\quad (i=1,2,3),
\ee
with the Wilson coefficients $C_{9,10}$ taken  at NNLL order 
(in the terminology of \rf{beneke:etal}). The ${\cal T}_{i}$ in \eq{subs:Tis}
are given by 
\bea
{\cal T}_1={\cal T}_\perp, \quad {\cal T}_2=\frac{2 E_{\kstar}}{m_B}{\cal T}_\perp,
 \quad {\cal T}_3={\cal T}_\perp+{\cal T}_\parallel,
\eea
where ${\cal T}_a$ ($a=\bot,\|$) contain factorizable (f) and non-factorizable (nf) 
contributions \cite{beneke:etal}:
\bea\label{nlodef}
{\cal T}_\perp &=& \xi_\perp(0)
\Bigg\{ C_\perp^{(0)} \frac{1}{(1-s/m_B^2)^2} +
\frac{\alpha_s}{3\pi} \Bigg[
\frac{C_\perp^{(1)}}{(1-s/m_B^2)^2}+ \kappa_\perp
  \lambda^{-1}_{B,+}\int_0^1 du\,\Phi_{K^*, \perp}(u)\nnu\\
&\times& [T_{\perp,+}^{(\mathrm{f})}(u)+
T_{\perp,+}^{(\mathrm{nf})}(u)] \Bigg] \Bigg\},
\eea
\bea\label{taus}
{\cal T}_\parallel &=& \xi_\parallel(0) \Bigg\{
C_\parallel^{(0)} \frac{1}{ {(1-s/m_B^2)^3}}
+\kappa_\| \frac{m_{K^*}}{E_{\kstar}} \lambda^{-1}_{B,-}(s)
\int_0^1 du\,\Phi_{K^*,\parallel}(u)\hat{T}_{\parallel,-}^{(0)}(u)\nnu\\
&+& \frac{\alpha_s }{3\pi} \Bigg[
\frac{C_\parallel^{(1)}}{(1-s/m_B^2)^3}  + \kappa_\| \frac{m_{K^*}}{E_{\kstar}} 
\Bigg(\lambda^{-1}_{B,+}\int_0^1 du\,\Phi_{K^*,\parallel}(u)[T_{\parallel,+}^{(\mathrm{f})}(u)+
T_{\parallel,+}^{(\mathrm{nf})}(u)]+
\lambda^{-1}_{B,-}(s)\nnu\\
&\times&\int_0^1 du\,\Phi_{K^*,\parallel}(u)\hat{T}_{\parallel,-}^{(\mathrm{nf})}(u)\Bigg) \Bigg] \Bigg\}, 
\eea
with 
\be
\kappa_a= \frac{\pi^2 f_B f_{K^*,a}}{3 m_B \xi_a(0)}, \quad
\hat{T}_{\parallel,\,-}^{(0,\mathrm{nf})}(u)
= \frac{(m_B \omega -s-i \epsilon)}{m_B \omega} {T}_{\parallel,\,-}^{(0,\mathrm{nf})}(u,\omega).
\ee 
Explicit expressions for the $T$'s, ${C}$'s, $\lambda^{-1}_{B,-}(s)$ and the light-cone 
wave functions  $\Phi_{K^* a}$, together with the remaining input parameters, 
can be found in \rf{beneke:etal}. 

Let us now analyse the impact of the NLL corrections on the
observables introduced in the preceding subsection. In particular, we are 
interested in the sensitivity of these quantities to a variation of the 
theoretical input parameters, which we take from Table 
$2$ of \cite{beneke:etal}. 

We proceed by computing the quantities defined in 
\eqs{def:asymmetries}{def:frac:pol} to NLL accuracy 
by replacing the Wilson coefficients according to \eq{subs:Tis}. 
We find that the dependence on the actual values of 
the soft form factors $\xi_{\bot,\|} (0)$ 
plays an important role in certain observables, as can be seen from 
Eqs.~(\ref{nlodef}) and (\ref{taus}). Furthermore,
we have explored the sensitivity of the NLL result to the
scale dependence of the observables, which is mainly due to the
hard-scattering correction, by changing the scale $\mu$ from
$m_b/2$ to $2 m_b$. This scale dependence is, apart from the error on the soft form
factors at $s=0$, the main source of uncertainty. We have
included the errors  associated with the input parameters in quadrature. 
Concerning a variation of $m_c/m_b$, which  affects 
mainly the contributions to the matrix element of the
chromomagnetic operator (specifically the functions 
$F^{(7,9)}_{1,2}$ in  \cite{NNLO:ME:formulae}),
we have chosen the range $0.27\leqslant m_c/m_b \leqslant 0.31$ 
\cite{NNLO:ME:formulae}, together with the 3-loop running for the strong 
coupling constant.

\begin{figure}
\begin{center}
\includegraphics[scale=1]{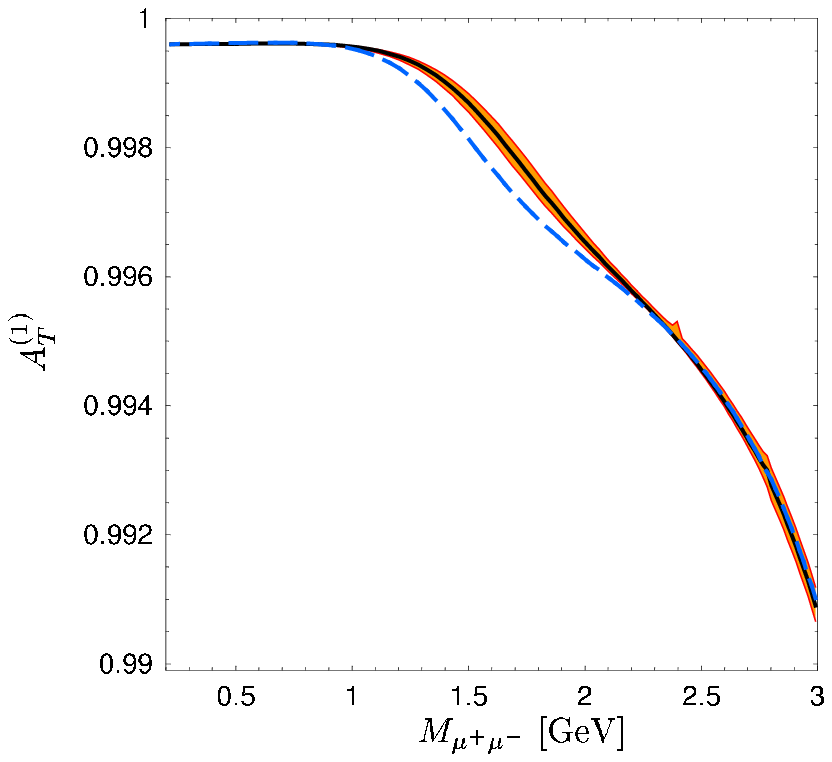}
\hspace{0.2em}
\includegraphics[scale=1]{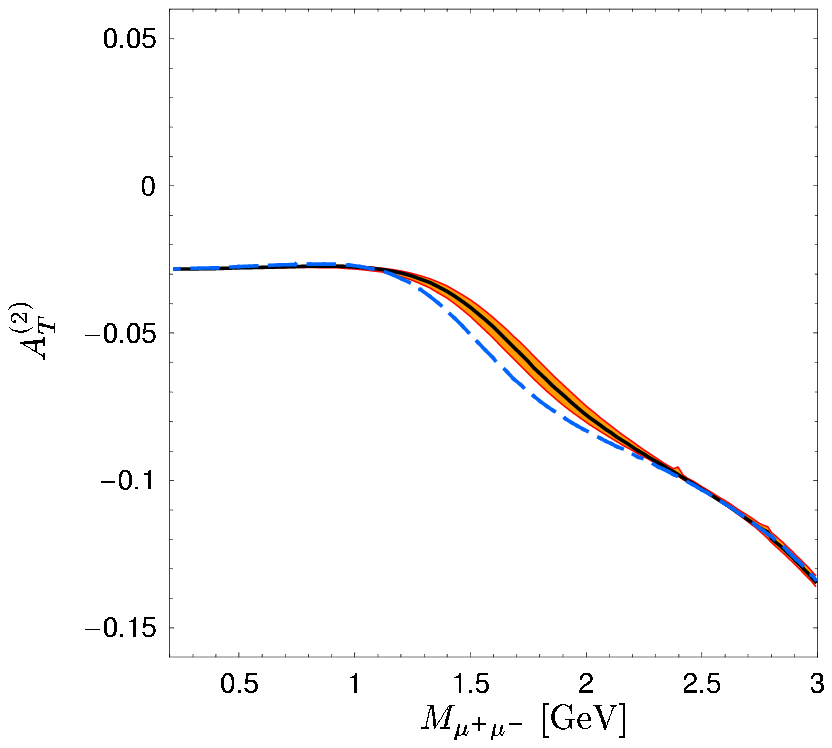}
\caption{SM predictions for the asymmetries $A_T^{(1)}$  and
$A_T^{(2)}$ as a function of the dimuon mass at
LL (dashed line) and NLL (solid line). The shaded area has been
obtained by varying the renormalization scale, the $b$-quark mass,  $m_c/m_b$, 
the input parameters according to \rf{beneke:etal}, and 
$\xi_\perp(0)$ as described in the text.}\label{SM:AT1:AT2}
\end{center}
\end{figure}

\begin{figure}[t]
\begin{center}
\includegraphics[scale=1]{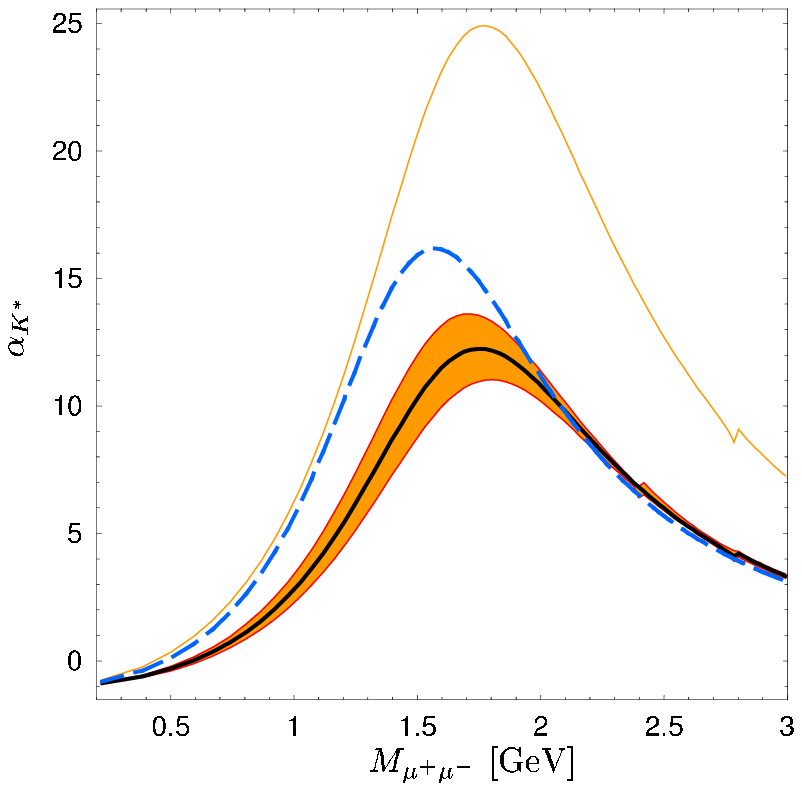}
\hspace{0.2em}
\includegraphics[scale=1]{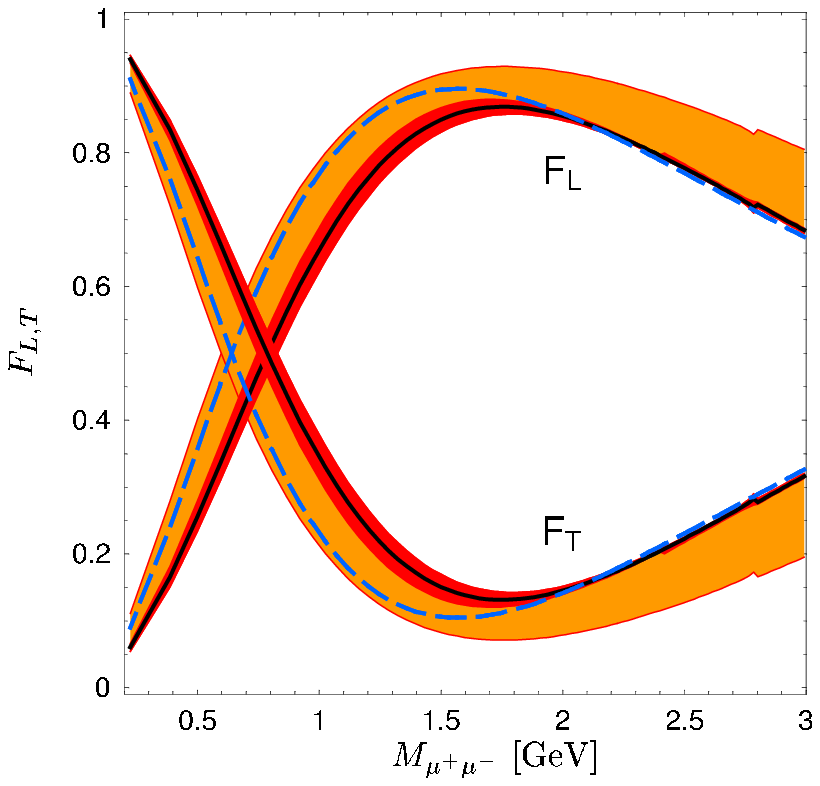}
\caption{\emph{Left plot:} The polarization parameter  $\alpha_{K^*}$ at LL
(dashed curve) and NLL (lower solid curve), as a function of the dimuon mass. 
The shaded area has been obtained
by varying  the theoretical input parameters for fixed 
$\xi_\perp(0)=0.35$ 
(see the text for details). To show the strong impact of 
$\xi_\perp(0)$, we also display the NLL result for  
$\xi_\perp(0)=0.24$ (upper solid curve). \emph{Right plot:} The SM prediction for the 
$K^*$ polarization fractions, adopting the same conventions as before and allowing for 
a variation of $\xi_\perp(0)$ between $0.24$ and $0.35$.
The inner dark region has been obtained by varying the theoretical input parameters for 
fixed $\xi_\perp(0)=0.35$.}
\label{SM:AL:FLT}
\end{center}
\end{figure}

Our results are summarized in  Figs.~\ref{SM:AT1:AT2} and  \ref{SM:AL:FLT}.~As 
expected, the transverse asymmetries $A^{(1)}_{T}({s})$ and $A^{(2)}_{T}({s})$ are 
the most promising observables. Indeed, from \eqs{LEL:tranversity:perp}{LEL:tranversity:zero} it is clear that the dependence on the hadronic form factors drops out in
the asymmetries at leading order in $1/m_b$ and $\alpha_s$. In this case, 
neglecting terms of 
$O(m_{K^*}^2/m_B^2)$, the SM predicts 
$A^{(1)}_{T}({s})\sim 1$ and
$A^{(2)}_{T}({s})\sim 0$. At NLL order, 
the impact of the soft form factors depends
on the relative size of the NLL corrections compared to the leading
order ones. In \fig{SM:AT1:AT2}, we plot $A^{(1,2)}_{T}$ versus 
the dimuon mass $M_{\mu^+\mu^-}$ including NLL order corrections 
\cite{beneke:feldmann,beneke:etal}
 to the form factor relations in
Eqs.~(\ref{form:factor:relations:LEL}).
%
%
Note that the small cusp, e.g., in the left plot of \fig{SM:AT1:AT2} is due 
to a variation of $m_c$ and it is an artifact of 
the $c{\bar c}$ threshold in the charm loop diagrams.
It is remarkable that there is only a small difference between the NLL result (solid
curve) and the LL one (dashed curve). Furthermore, the sensitivity of the transverse 
asymmetries to the theoretical input parameters is rather weak.
Indeed, neither the scale dependence nor  the inclusion of the
hadronic uncertainties (added in quadrature) induces large
deviations. 
We stress that the shaded areas in 
\fig{SM:AT1:AT2} also take into account  the 
possibility of a much smaller value for the soft form factor, namely
$\xi_\perp(0)=0.24$, which is favoured  by a fit to
experimental data on ${\mathcal B}(B\rightarrow K^* \gamma)$ 
\cite{beneke:etal,ff:BKsg}. 
The small impact of the theoretical uncertainties found,  including 
the NLL corrections, shows the robustness of the transverse asymmetries 
and 
makes them
an ideal place to search for new physics.

%
%

The predictions for $\alpha_{K^*}$ and $F_{L,T}$ in the SM are
shown in \fig{SM:AL:FLT}.
As can be seen from the left plot in \fig{SM:AL:FLT}, 
there is a strong impact of the NLL corrections (solid curves)
on $\alpha_{K^*}$. Moreover, the error band is substantially
larger even for a fixed value of $\xi_\perp(0)=0.35$. A more detailed analysis of 
the dependence on the value of $\xi_\perp(0)$ shows that if we 
consider instead $\xi_{\perp}(0)=0.24$, which  corresponds
to the upper solid curve, 
the deviation would induce an even larger error band. In fact, as 
can be seen from the LL expressions  in 
Eqs.~(\ref{LEL:tranversity:perp})--(\ref{LEL:tranversity:zero}), together with 
(\ref{def:pol:param}) and (\ref{def:frac:pol}), there is no
 cancellation between the soft form factors. As a result, changing
$\xi_\perp(0)$ from $0.35$ to $0.24$ should enhance 
the maximum of $\alpha_{K^*}$, as a function of  the dimuon mass, 
by roughly a factor of two. This is also the case at NLL order, 
as can be inferred from the left plot in \fig{SM:AL:FLT}.
The strong sensitivity of $\alpha_{K^*}$ to the poorly known
quantity $\xi_\perp(0)$ therefore requires a better theoretical control on the 
soft form factor.

As far as $F_{L,T}$ are concerned, they also depend on $\xi_\perp(0)$, 
but the fact that the normalization factor includes  
$A_0$ reduces the impact on the soft form factor. 
The right  plot in \fig{SM:AL:FLT}   shows the LL result (dashed curve),
the NLL (solid curve) and two bands. The internal one includes all
errors for fixed $\xi_\perp(0)=0.35$, while the wider band 
includes a variation of $\xi_\perp(0)$ between $0.24$ and $0.35$.
Note that our results for the $K^*$
polarization fractions are slightly different from those of
\rf{ali:safir}. This can be traced to the different parametrization and normalization
of the soft form factors appearing in \eqs{LEL:tranversity:perp}{LEL:tranversity:t}
(see Fig.~1 in \cite{ali:safir}).

Finally, we have computed the integrated observables including NLL corrections  
and using $\xi_\perp(0)=0.35\pm 0.07$. Our results for  the low dimuon mass region
are listed  in Table \ref{SM:summary}. Notice that the theoretical 
uncertainties of the asymmetries ${\mathcal A}_T^{(1,2)}$  amount to less then 
$7\%$, so that its measurement will allow a precise test of the SM.
%
%
\begin{table}
\begin{center}
\caption{SM predictions for the observables defined in 
\eqs{def:asymmetries}{def:frac:pol} when integrated over the low dimuon mass region $2m_\mu \leqslant M_{\mu^+\mu^-} \leqslant
2.5\ \GeV $. 
The form factors $\xi_{\bot,\|}$ and the input parameters are taken from \rf{beneke:etal}. \label{SM:summary}}
\bigskip

\begin{tabular}{ccccc}\hline\hline
 ${\mathcal A}_T^{(1)}$& ${\mathcal A}_T^{(2)}$& $\mbox{\bm $\alpha$}_{K^*}$ &
${\mathcal F}_L$ & ${\mathcal F}_T$  \\
\hline \vspace{-1em}\\
$0.9986\pm 0.0002$ & $-0.043\pm 0.003$ & $3.47\pm 0.71$ & $0.69\pm 0.03$
& $0.31\pm 0.03$
\\\hline\hline\end{tabular}
\end{center}
\end{table}

\subsection{Impact of new physics on the \bm$K^*$ polarization}
We now study model-independently the implications  of right-handed currents for  
the observables defined in \eqs{def:asymmetries}{def:frac:pol}. Since the low dimuon 
mass region is dominated by the photon  pole, $|C_7^{\mathrm{eff}(\prime)}|^2/s$, 
we do not take into account the contributions of the chirality-flipped
operators $\O_{9,10}^\prime$,  but will allow for deviations of $C_{9,10}$ from their 
SM values.\footnote{The coefficient $C_9$ is related to the effective one introduced in \eq{matrix:ele} 
through $\ceff = C_9 + Y(s)$, where $Y(s)$ contains contributions from the four-quark operators
${\mathcal{O}}_{1-6}$ (see \rf{wilson:coeffs:SM} for details).}  
Examples of new-physics  scenarios that could give
sizable contributions to  ${C_7^{\mathrm{eff}}}^{(\prime)}$ include the 
left-right model \cite{LRmodel}, the unconstrained minimal supersymmetric standard 
\cite{Borzumati:1999qt}, and an $\mathrm{SO} (10)$ SUSY GUT model with large 
mixing between $\tilde{s}_R$ and $\tilde{b}_R$ \cite{Chang:2002mq}.

In our analysis we use the Wilson coefficients and soft form 
factors $\xi_{\bot,\|}$ of  \rf{beneke:etal}. 
Furthermore, we take, for simplicity, the leading-order condition
\be\label{bound} 
|{C_7^{\mathrm{eff}}}|^2 + |{C_7^{\mathrm{eff}}}^\prime|^2\leqslant 1.2
|{C_7^{\mathrm{eff}, \mathrm{SM}}}|^2, 
\ee
which follows from the requirement that the theoretical prediction for 
${\mathcal B}(B \to X_s \gamma)$ \cite{Gambino:2001ew} 
is within $2\sigma$ of the experimental average 
$(3.52^{+0.30}_{-0.28})\times 10^{-4}$ \cite{HFAG}. For the transversity 
amplitudes describing the polarization states of the $K^*$, we use the 
expressions given in \eqs{a_perp}{a_long}, together with the leading-order 
form factor relations in Eqs.~(\ref{form:factor:relations:LEL}).
We discuss the observables defined in \eqs{def:asymmetries}{def:frac:pol} in turn.

\begin{figure}[!t]
\begin{center}
\includegraphics[scale=0.53]{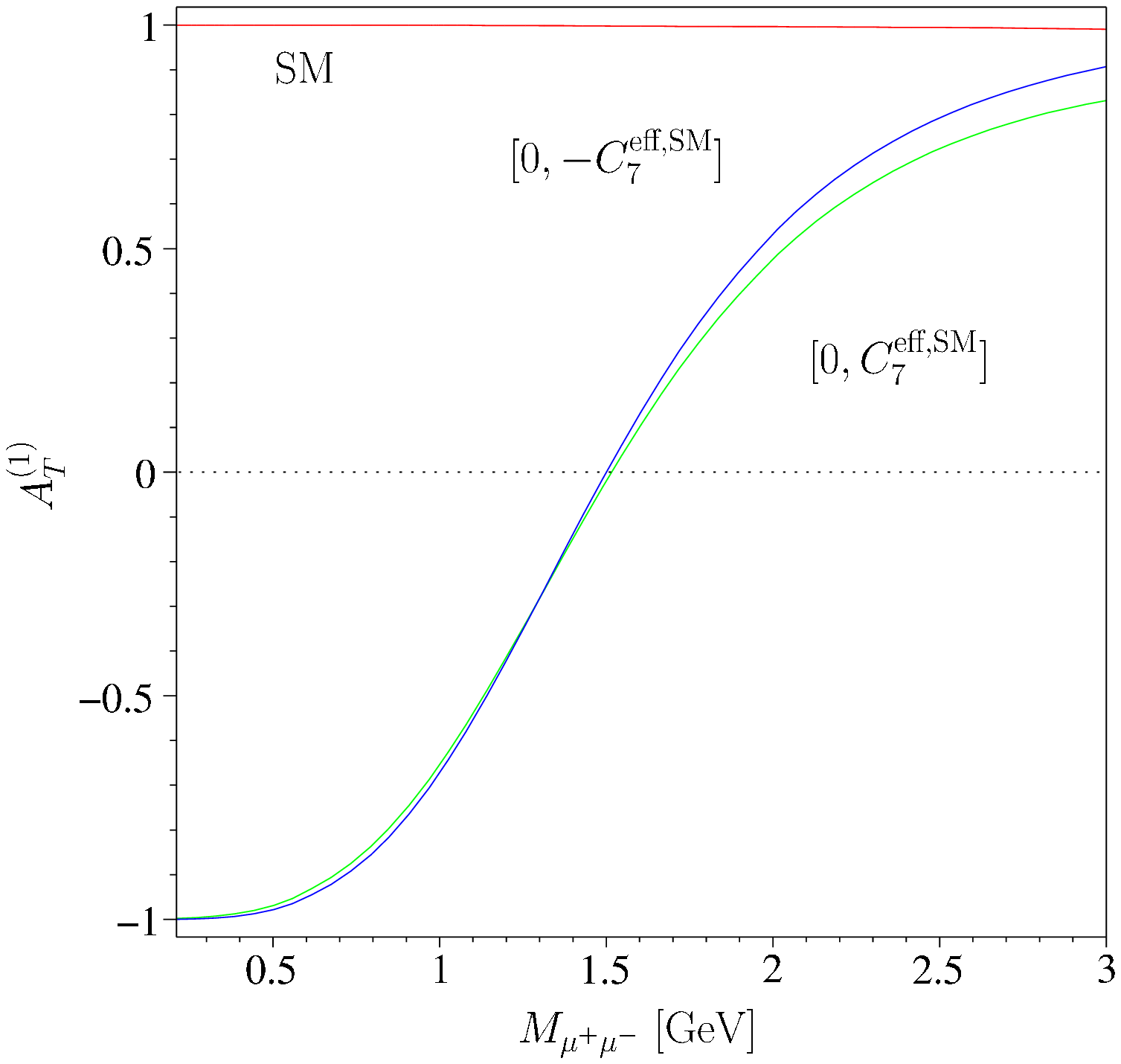}
\hspace{.5em}
\includegraphics[scale=0.53]{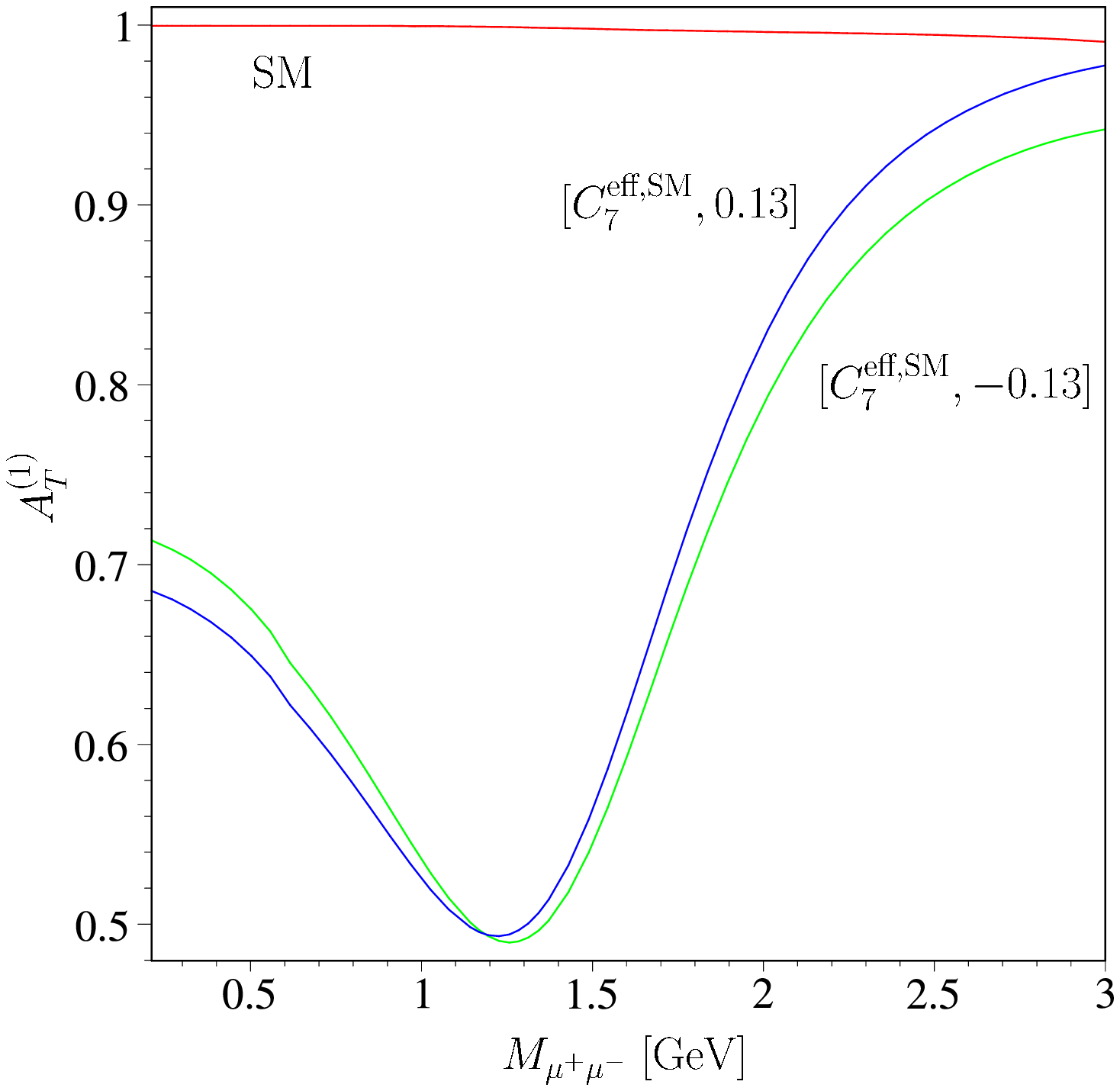} 
\caption{The transverse asymmetry $A^{(1)}_{T}$ as a function of the dimuon mass, for different choices of $[C_7^{\mathrm{eff}},C_7^{\mathrm{eff}\prime}]$. The 
SM prediction corresponds to the case $[C_7^{\mathrm{eff, SM}},0]$. We have taken 
into account the bound in \eq{bound} and the SM values for $C_{9,10}$  
from \rf{beneke:etal}.}\label{fig:AT1}
\end{center}
\end{figure}

\begin{figure}[!ht]
\begin{center}
\includegraphics[scale=0.53]{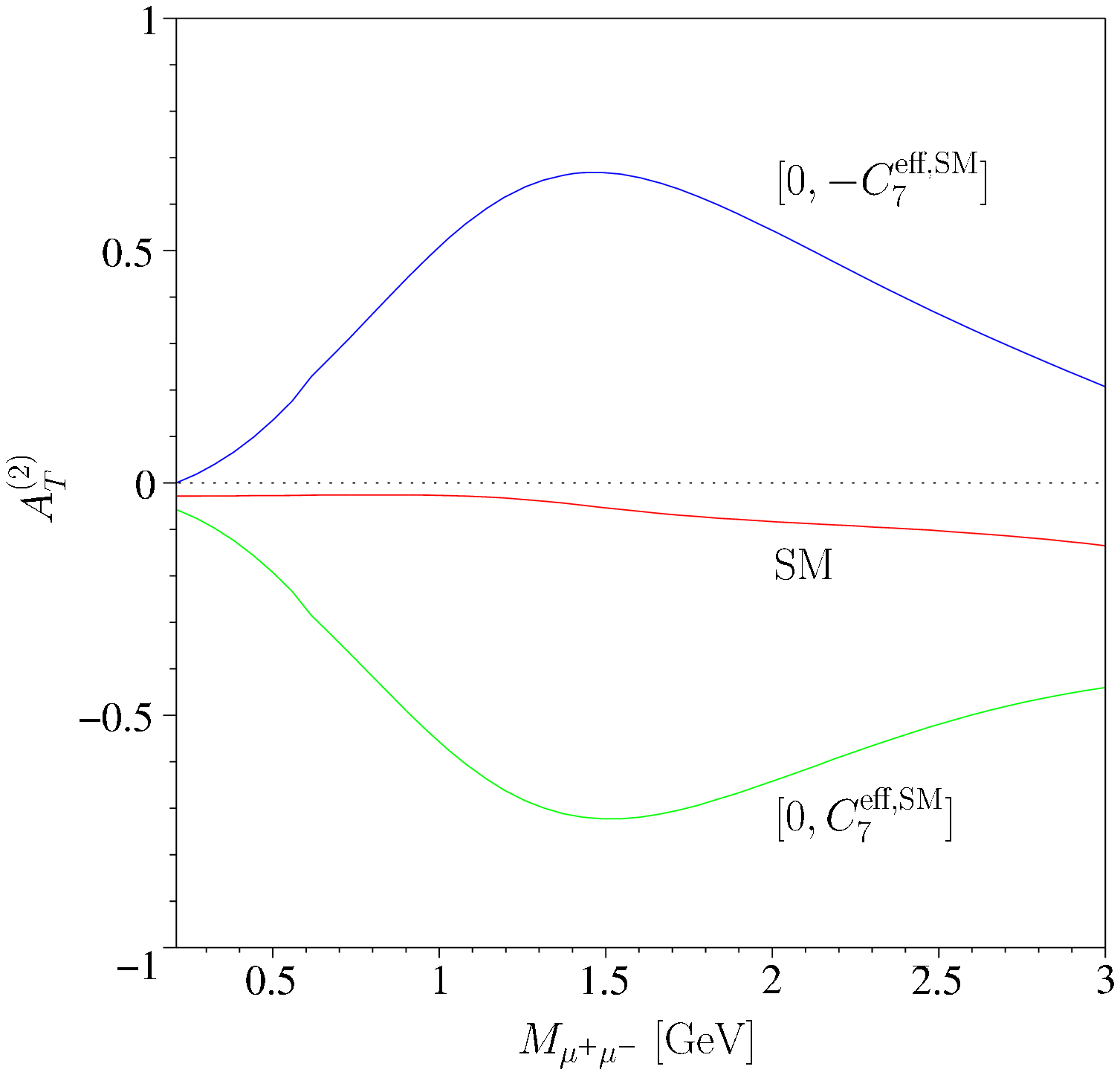}
\hspace{.5em}
\includegraphics[scale=0.53]{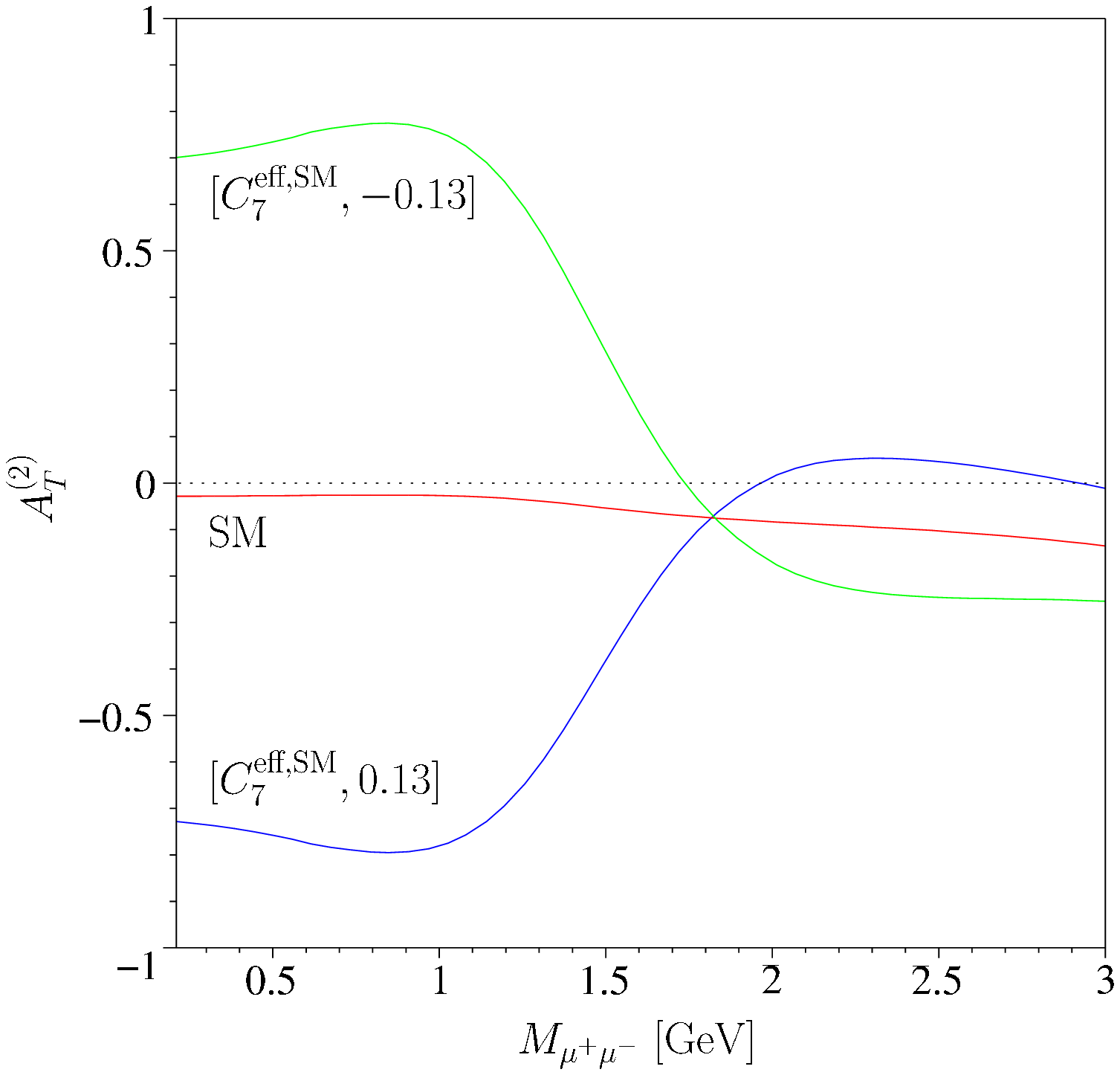} 
\caption{The asymmetry $A^{(2)}_{T}$ as a function of  the dimuon mass, for different new-physics scenarios. See \fig{fig:AT1} for details.}\label{fig:AT2}
\end{center}
\end{figure}

(i) $A^{(1,2)}_{T}$. The corresponding distributions as a 
function of the dimuon mass are 
shown in  Figs.~\ref{fig:AT1} and \ref{fig:AT2} for different sets of 
$[C_7^{\mathrm{eff}},C_7^{\mathrm{eff}\prime}]$. 
The new-physics contributions to $C_7^{\mathrm{eff} (\prime)}$ are chosen such that 
the effects are striking, while being consistent 
with the $b\to s \gamma$ bound in \eq{bound}. 
As can be seen, the impact  of right-handed currents on  $A^{(1)}_{T}$ in the low 
dilepton mass region is maximal for the 
extreme case where new physics cancels the SM contributions to 
$C_7^{\mathrm{eff}}$, so that $A^{(1)}_{T}=   
-{\hat N}\;[\; |C_7^{\mathrm{eff}\prime}|^2/s^2+{O}(
C_7^{\mathrm{eff}\prime} C_{9,10})/s+{O}(C_{9,10}^2)]/({d\Gamma/ds})$, where
${\hat N}= 16 |N|^2 m_b^2 \lambda^{1/2} (m_B^2-m_{K^*}^2)$. 
In this case, $A^{(1)}_{T}$ has a zero in the presence of new physics, 
contrary to the SM case. (Our results for the new-physics contributions  
in the left plot of \fig{fig:AT1} are very similar to the ones obtained in the 
left-right model of \cite{dmitri}.) But even if there is only a small contribution from  
right-handed currents to the transverse polarization of $K^*$ 
(right plot in \fig{fig:AT1}), the effects are quite different from the SM predictions. 
As far as $A^{(2)}_{T}$ is concerned (\fig{fig:AT2}), it is not only sensitive to the magnitude of the 
right-handed current coupling $C_7^{\mathrm{eff}\prime}$, but also to its sign. 
Moreover, for very low dilepton masses, 
$A^{(2)}_{T}$ can be used to determine the helicity amplitudes in the radiative 
$B\to K^* \gamma$ decay, as was shown in \rf{BtoKpill:gamma}.
We note parenthetically that this asymmetry was studied in \rf{CS:etal}
within a generic supersymmetric extension of 
the SM, but without using the soft form factor relations in
Eqs.~(\ref{form:factor:relations:LEL}).  

%
%
%
%
%
\begin{figure}
\begin{center}
\includegraphics[scale=0.53]{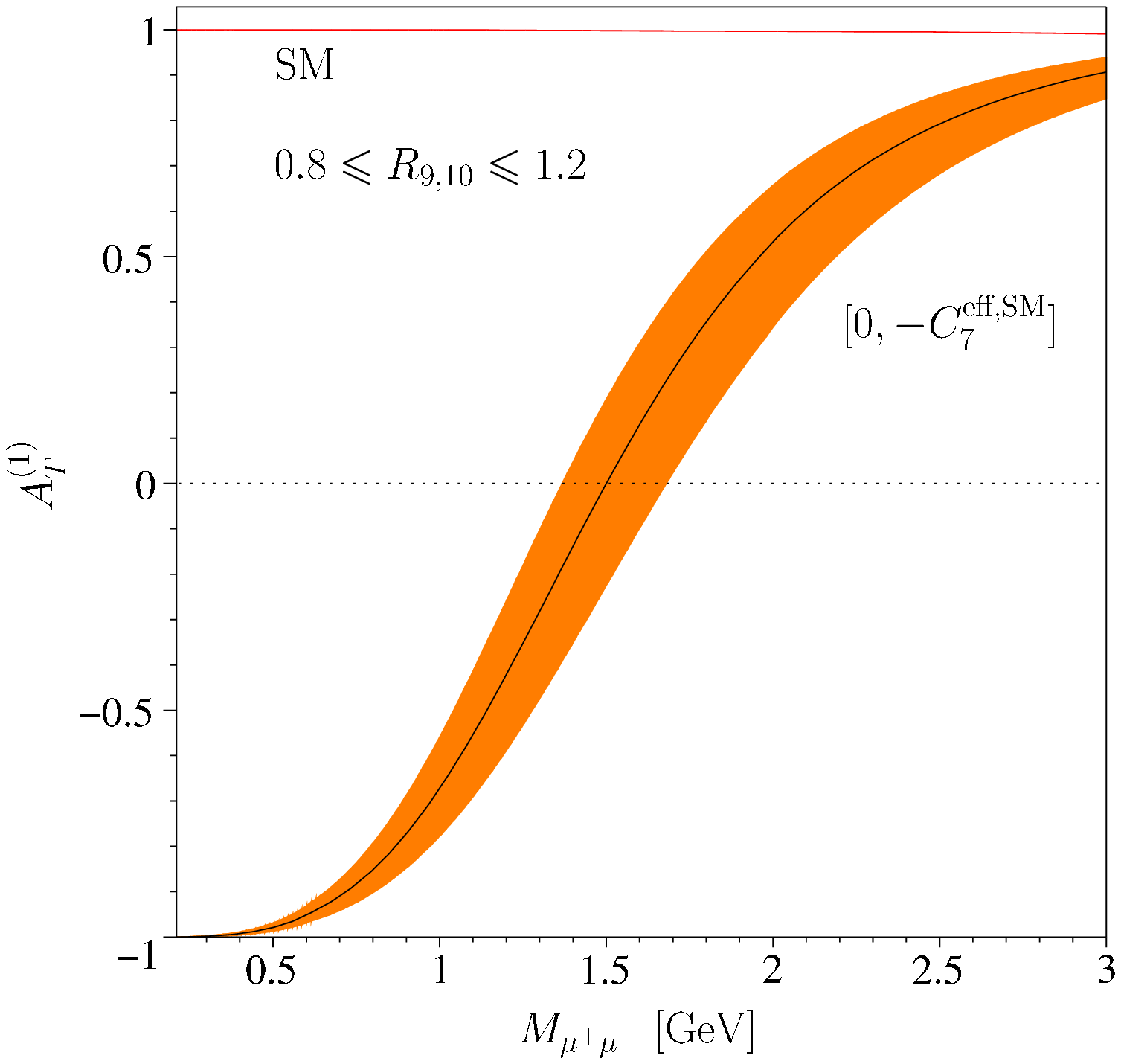}
\hspace{0.5em}
\includegraphics[scale=0.53]{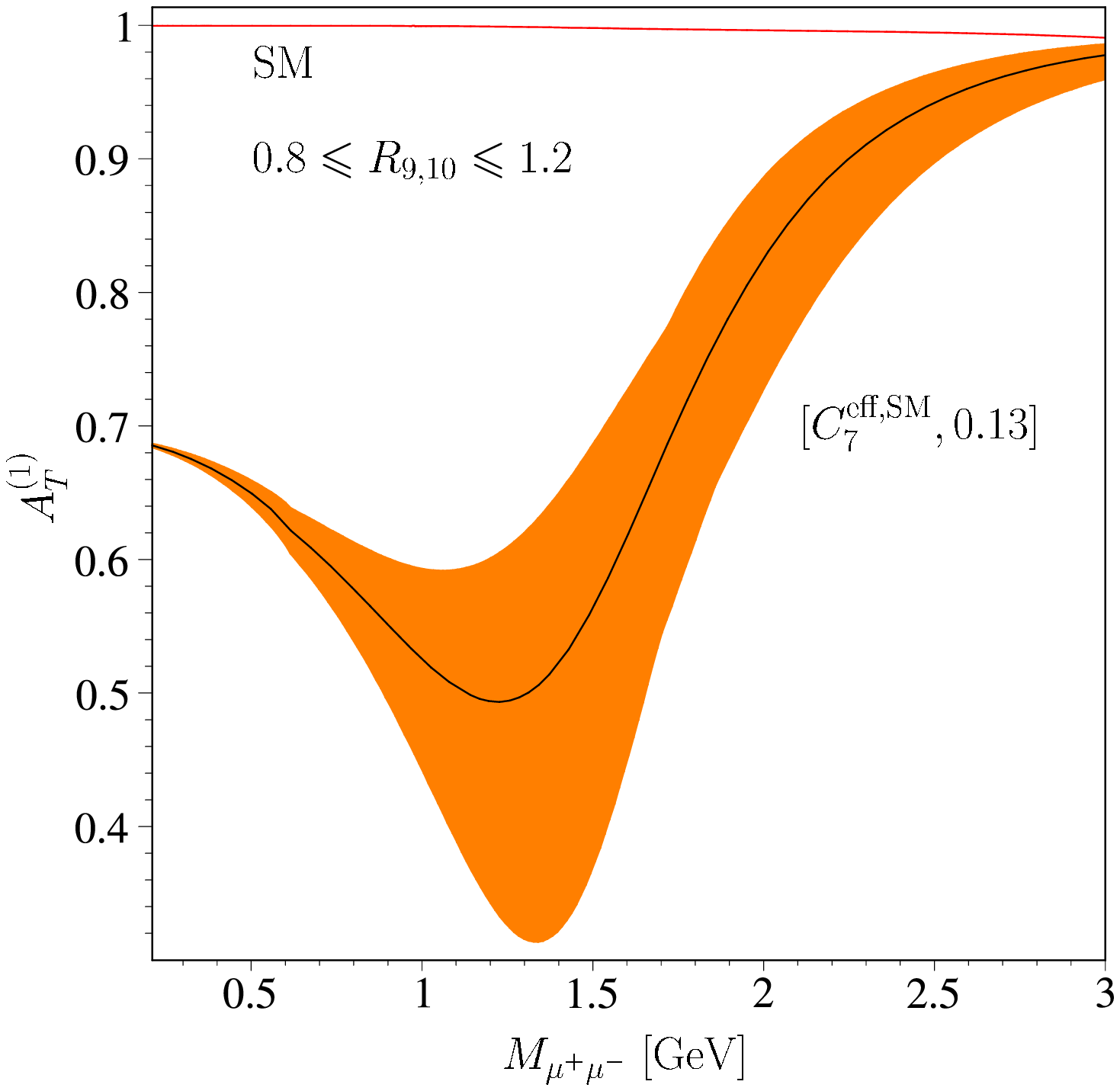}
\caption{The asymmetry $A_T^{(1)}$ vs the dimuon mass  for fixed $[\cseff,\cseffP]$,
including new-physics contributions to $C_{9,10}$. The shaded areas have been 
obtained by varying $R_{9,10} \equiv C_{9,10}/C_{9,10}^{\mathrm{ SM}}$ in a 
range that is consistent with present data on rare $B$ decays 
(see, e.g., \rfs{Ali:2002jg,Hiller:2003js}). The inner solid lines correspond to the 
case where $C_{9,10}= C_{9,10}^{\mathrm{SM}}$.  
\label{fig:AT1_C910var}}
\end{center}
\end{figure}
%
%
\begin{figure}
\begin{center}
\includegraphics[scale=0.53]{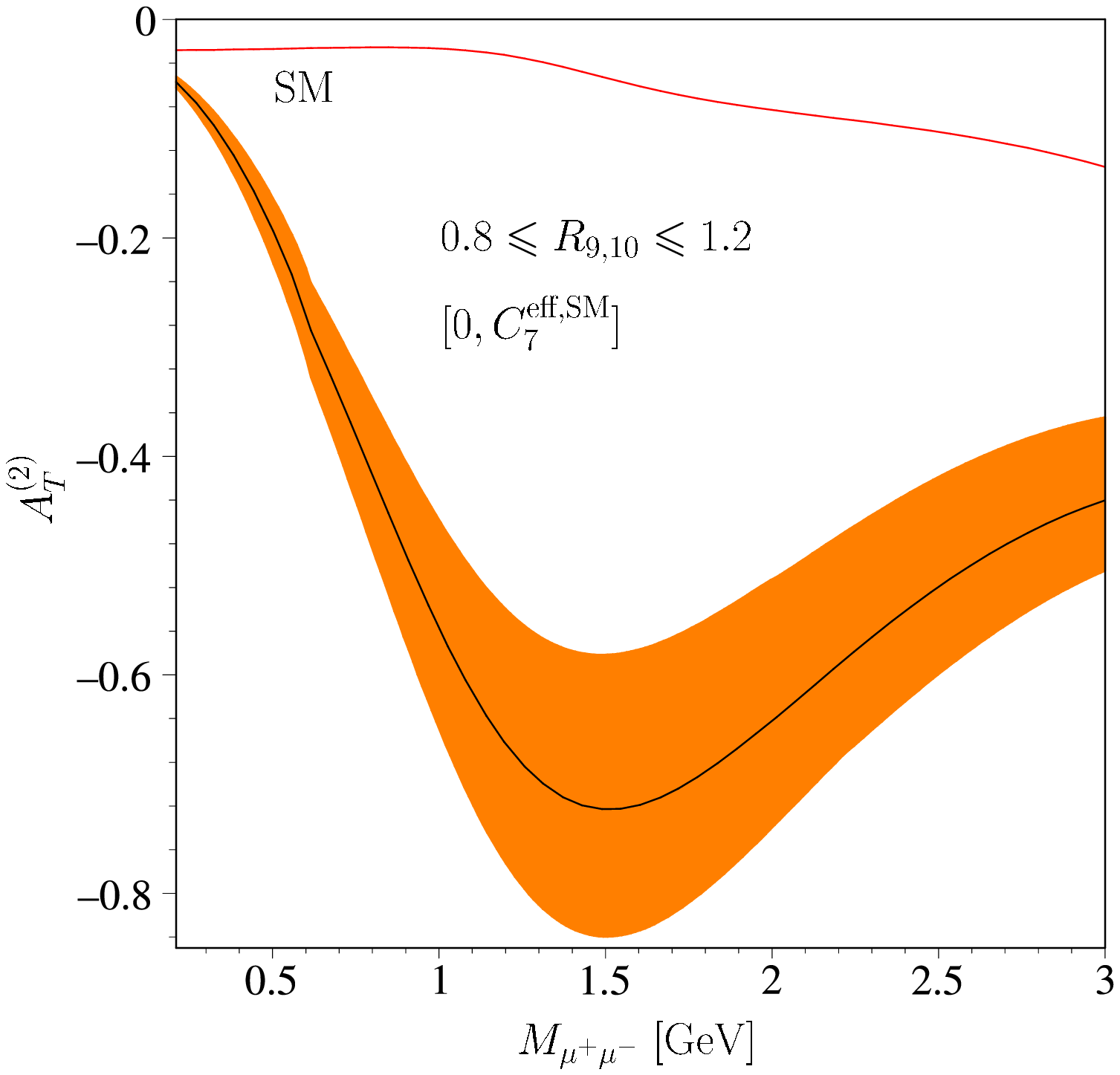}
\hspace{0.5em}
\includegraphics[scale=0.53]{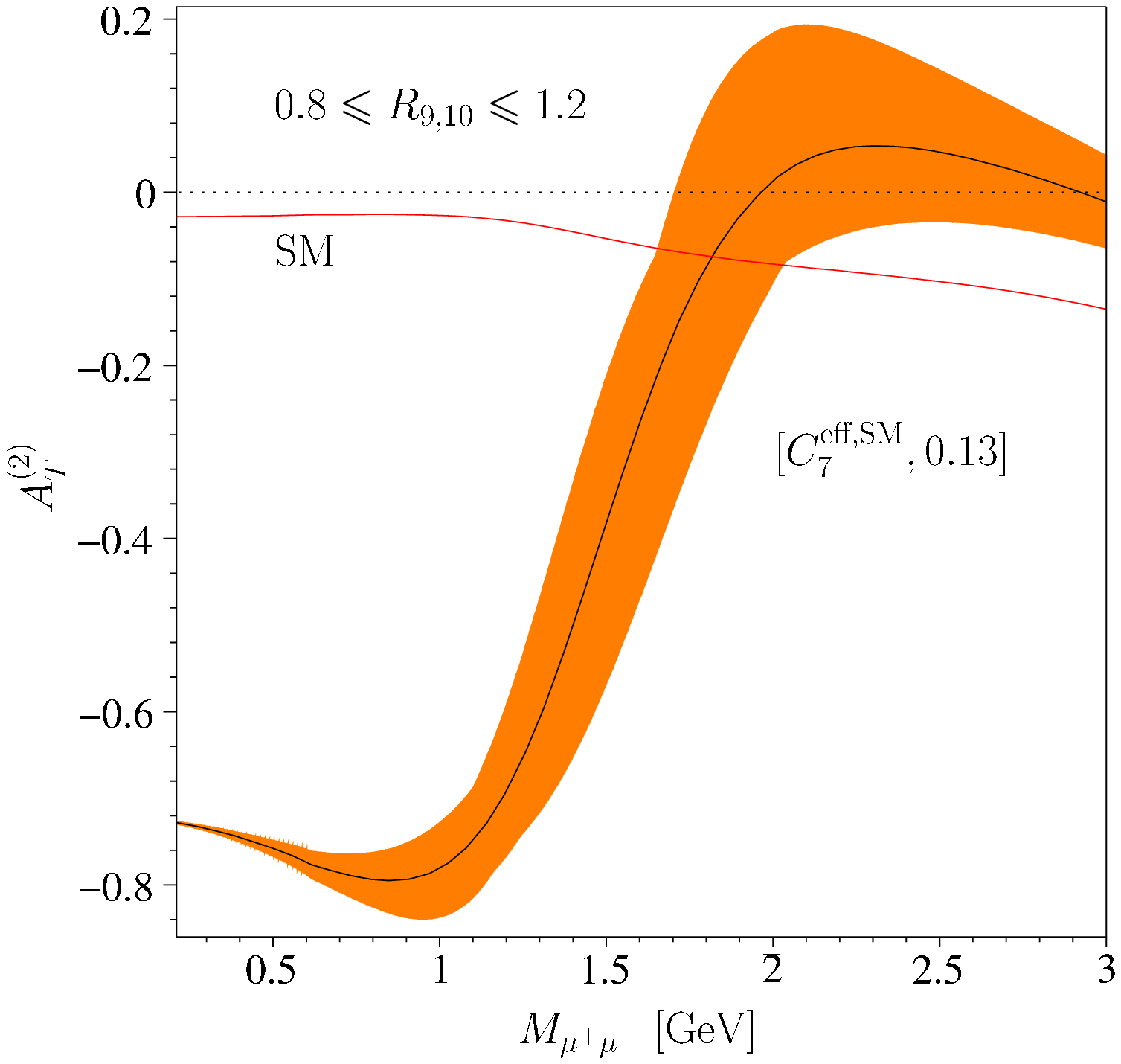}
\caption{$A^{(2)}_{T}$ vs the dimuon mass in the presence of new-physics 
contributions to $C_7^{\mathrm{eff}(\prime)},C_{9,10}$. See \fig{fig:AT1_C910var} for details.
\label{fig:AT2_C910var}}
\end{center}
\end{figure}

So far we have assumed that new physics enters only via
$C_7^{\mathrm{eff}(\prime)}$ while the remaining coefficients are SM-like.
Allowing for non-standard contributions to the coefficients
$C_{9,10}$, and taking into account the constraints from rare $B$ decays
\cite{Ali:2002jg,Hiller:2003js} (see also \cite{Gambino:2004mv}),
we obtain the asymmetries shown
in Figs.~\ref{fig:AT1_C910var} and \ref{fig:AT2_C910var}. As can be seen,
a determination of the magnitude of the right-handed currents is possible even in
the presence of new-physics contributions to $C_{9,10}$ of the order of $20\%$.
In view of this and the small theoretical uncertainties expected from our previous
discussion in the SM, it is clear that
the transverse asymmetries $A_T^{(1,2)}$ can be an especially
useful probe of the electromagnetic penguin operator
${\mathcal{O}}_7^\prime$. We therefore conclude that a measurement
of $A^{(1,2)}_{T}$ in the low dilepton mass region different from their SM values
could be a hint of new physics with right-handed quark currents.

\begin{figure}
\begin{center}
\includegraphics[scale=0.53]{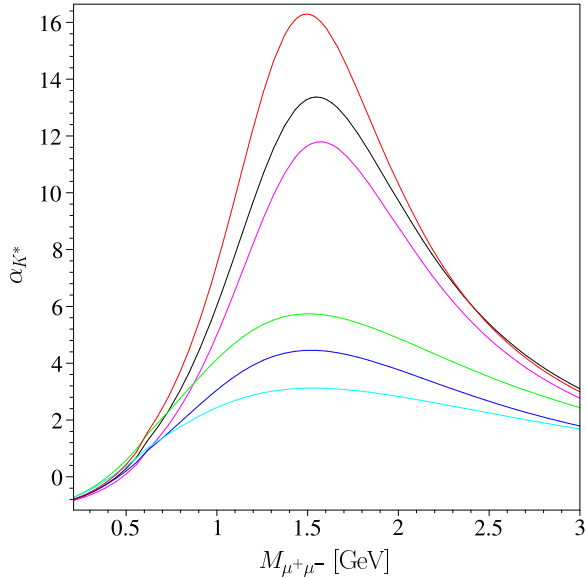}
\end{center}
\caption{The $K^*$ polarization parameter $\alpha_{K^*}$ as a function of the 
dimuon mass for different sets  of $[C_7^{\mathrm{eff}},C_7^{\mathrm{eff}\prime}]$ (from top to bottom): 
(i) $[C_7^{\mathrm{eff,SM}},0]$ (SM), 
(ii) $[C_7^{\mathrm{eff,SM}},-0.13]$, 
(iii) $[C_7^{\mathrm{eff,SM}},0.13]$,
(iv) $[0,C_7^{\mathrm{eff,SM}}]$, 
(v) $[0,-C_7^{\mathrm{eff,SM}}]$, 
(vi) $[-C_7^{\mathrm{eff,SM}},0]$. 
The remaining Wilson coefficients are assumed to be SM-valued.}
\label{fig:alpha}
\end{figure}

(ii) $\alpha_{K^*}$. Our results for the $K^*$ polarization parameter are  shown in 
\fig{fig:alpha}. Like in the case of the transverse asymmetries, 
new physics can give large contributions, but only for extreme scenarios. 
We emphasize that some of the new-physics scenarios shown in 
\fig{fig:alpha} are indistinguishable if we also allow $C_{9,10}$ to deviate from 
their SM values. Moreover, from our discussion of the SM prediction, we expect 
important NLL corrections to $\alpha_{K^*}$ in the presence of new 
physics and a large theoretical error due to  the poorly known form factor
$\xi_\perp(0)$. 
%
%

\begin{figure}
\begin{center}
\includegraphics[scale=0.53]{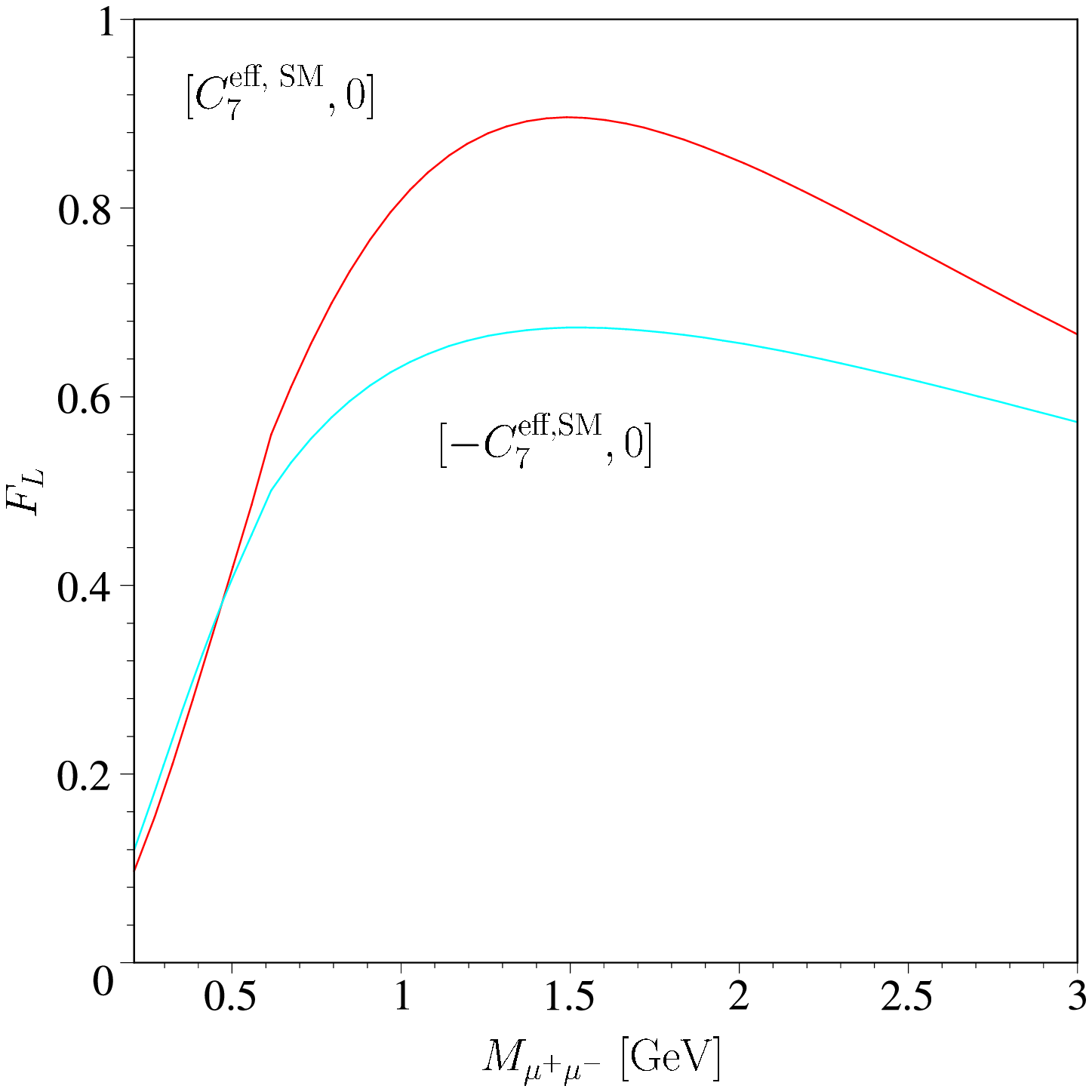}
\hspace{0.5em}
\includegraphics[scale=0.53]{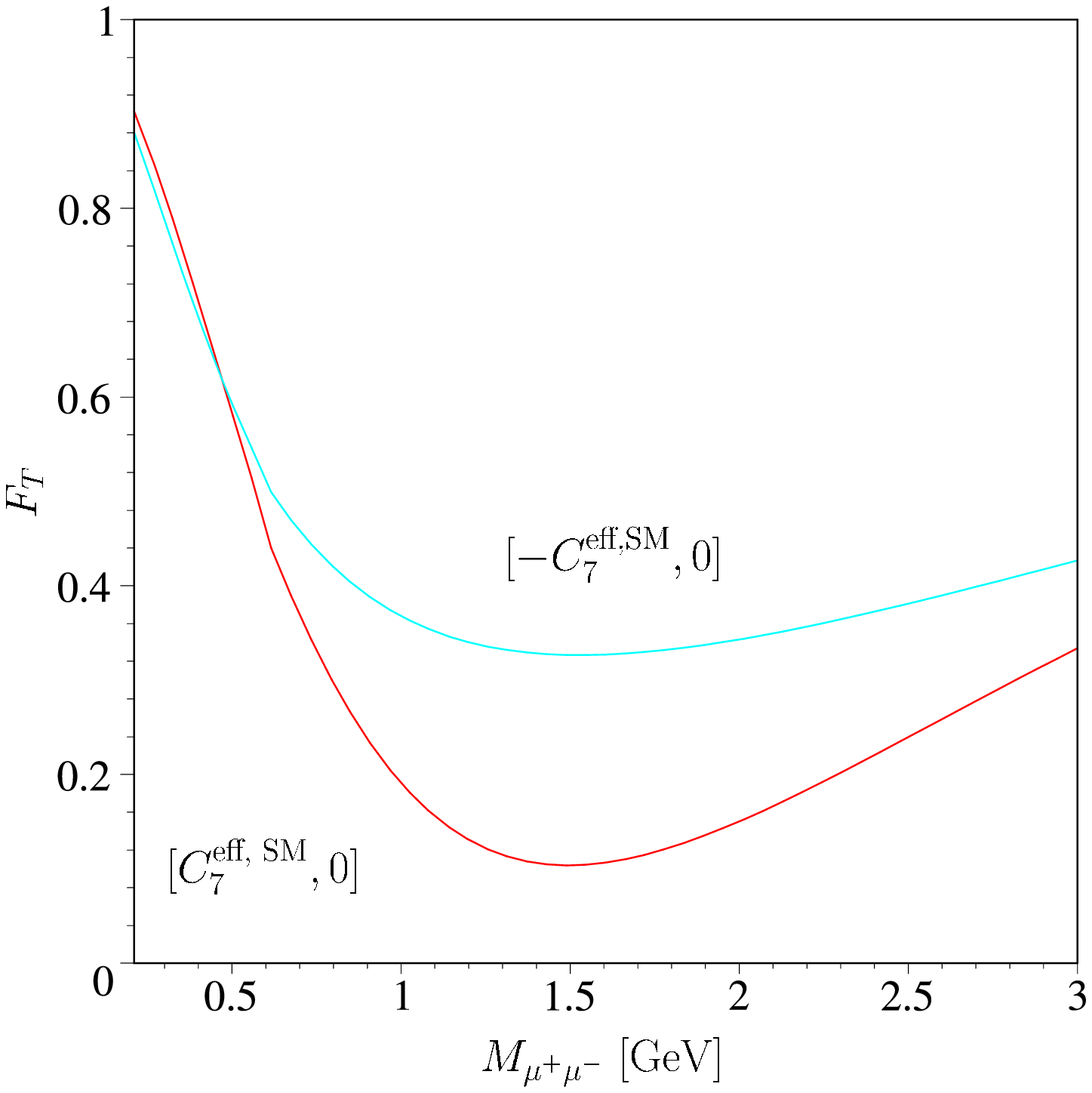}
\end{center}
\caption{Longitudinal and transverse $K^*$ polarization fractions vs   
the dimuon mass, for two scenarios of 
$[C_7^{\mathrm{eff}},C_7^{\mathrm{eff}\prime}]$ and $C_{9,10}$  
being SM-like. Since the remaining new-physics cases displayed in \fig{fig:alpha} 
lie between these two curves, they are not shown here.}
\label{fig:ft:fl}
\end{figure}

(iii) $F_{L,T}$. The longitudinal and transverse polarization fractions are plotted 
in \fig{fig:ft:fl} for two scenarios of $C_7^{\mathrm{eff} (\prime)}$. 
Recalling that the polarization fractions are related to the 
$K^*$ polarization parameter, $\alpha_{K^*}=2F_L/F_T-1$, the conclusions drawn 
in (ii) also apply to $F_{L,T}$ except that the uncertainty induced by 
the parameter $\xi_\perp(0)$ is  smaller. As far as  additional 
operators are concerned, we merely mention that their impact on $F_L/F_T$ was  
studied model-independently in \rf{Aliev:1999gp}, but for $\cseffP=0$ and 
without using the soft form factor relations in Eqs.~(\ref{form:factor:relations:LEL}).
%
%

\newsection{Summary and conclusions}\label{summary}
In this paper we have performed a detailed analysis of the $K^*$  polarization states 
in the decay ${B}^0 \to K^{*0}(\to K^- \pi ^+)l^+l^-$. We have focused on the kinematic 
region in which the energy of the $K^*$ is of order $O(m_b)$, so that 
the theoretical predictions for the heavy-to-light form factors 
 are applicable \cite{beneke:feldmann,beneke:etal}. Within 
the framework of the SM, we have taken into account 
factorizable as well as  non-factorizable corrections at NLL order. We 
have shown that the asymmetries $A_T^{(1,2)}$, which involve 
transversely polarized $K^*$, are largely free of hadronic uncertainties. 
At leading order in the heavy quark and large energy 
expansion, we have found that the dependence of the transverse  asymmetries on 
the hadronic form factors completely drops out.~Taking into account 
NLL corrections 
of order  $\alpha_s$ \cite{beneke:etal}, and varying the 
theoretical input parameters, 
the error on the integrated transverse asymmetries in the low 
dimuon mass region is found to be less than $7\%$.
Within the SM, we obtain at NLL order ${\mathcal A}_T^{(1)}=0.9986\pm 0.0002$ and  
${\mathcal A}_T^{(2)}=-0.043\pm 0.003$. Thus, a measurement of  
$A_T^{(1,2)}$ will allow  a precise test of the SM. 
We have also investigated the $K^*$ polarization fractions in 
the low dimuon mass region. At NLL order, the   
longitudinal and transverse polarization fractions are predicted to be  
$(69\pm 3)\%$ and $(31\pm 3)\%$ respectively, and hence 
$\Gamma_L/\Gamma_T = 2.23\pm 0.31$.

We further studied model-independently the implications of new physics for the 
$K^*$ polarization states. Since the low dilepton mass 
region is dominated by the photon pole, 
$|C_7^{\mathrm{eff}(\prime)}|^2/s$, we have first concentrated 
on such new-physics scenarios
that can give appreciable contributions to the Wilson coefficients 
$C_7^{\mathrm{eff}(\prime)}$ of the electromagnetic dipole operators 
${\mathcal{O}}_7^{(\prime)}\sim m_b (\bar{s} \sigma_{\mu \nu} P_{L(R)} b )F^{\mu \nu}$. 
Taking into account constraints on the inclusive $b\to s \gamma$ branching ratio and assuming $C_{9,10}$ being SM-like, we have found large effects on the 
transverse asymmetries $A_T^{(1,2)}$ (Figs.~\ref{fig:AT1} and \ref{fig:AT2}) and on 
the $K^*$ polarization parameter 
$\alpha_{K^*}$ (\fig{fig:alpha}). While the former observables provide a useful 
tool to search for new physics,  the latter still suffers from theoretical 
uncertainties due to the soft form factor $\xi_\perp(0)$ (see \fig{SM:AL:FLT}). 
Nevertheless,   the $K^*$ polarization parameter will provide  
valuable information on non-standard physics once we have better control on
the actual value of $\xi_\perp(0)$. 
We have also investigated the 
implications of new physics for the longitudinal and transverse polarization fractions 
(\fig{fig:ft:fl}) whose measurement will allow to constrain beyond-the-SM  scenarios.

In addition to the aforementioned scenario with $C_{9,10}$ being SM-valued, 
we have investigated the case where these coefficients receive 
additional contributions. Focusing on the transverse asymmetries $A_T^{(1,2)}$, and 
taking into account  experimental data on rare $B$ decays, we have found that 
$A_T^{(1,2)}$  are still sensitive to the electromagnetic dipole operator 
${\mathcal{O}}_7^{\prime}$ (Figs.~\ref{fig:AT1_C910var} and \ref{fig:AT2_C910var}).
Thus, they provide an especially useful tool to search for  
right-handed currents in the low dilepton mass region. 

To sum up, the study of the angular distribution of the decay  
${B}^0 \to K^{*0}(\to K^- \pi ^+)l^+l^-$ provides valuable information on 
the $K^*$ spin amplitudes.~This enables us to probe non-standard 
interactions in a way that is not possible through measurements of 
the branching ratio and the lepton forward-backward asymmetry. 
Of particular interest is the lower part of the dilepton invariant mass region, 
where the hadronic uncertainties can be considerably reduced by exploiting 
the heavy-to-light form factor relations in the heavy quark and large-$E_{K^*}$ limit.

\section*{Acknowledgments}
We would like to thank Thorsten Feldmann for useful correspondence and 
comments. We are also grateful  to  
Gudrun Hiller for her comments on the manuscript. 
F.K.~would like to thank the theory group at IFAE for their 
support and kind hospitality. J.M.~acknowledges financial support 
from the Ramon y Cajal Program and FPA2002-00748.
The research of F.K. was partially supported by the Deutsche
Forschungsgemeinschaft under contract Bu.706/1-2. 

\begin{figure}
\begin{center}
\includegraphics[scale=0.9]{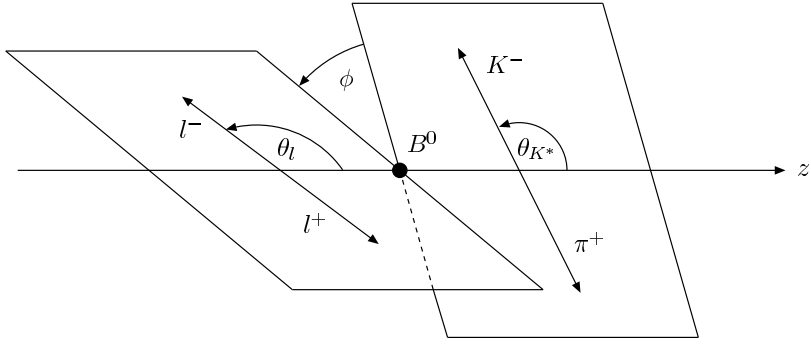}
\end{center}
\caption[]{Definition of kinematic variables in the decay 
${B^0}\to K^{*0}(\to K^- \pi ^+)l^+l^-$.}\label{fig:kinematics}
\end{figure}

\appendix
\newsection{Angular distribution of  \bm
${B^0}\to K^{*0}(\to K^- \pi ^+)l^+l^-$}\label{kin}
In this appendix we give the differential decay rate formula for finite lepton mass.
Assuming the $K^*$ to be on the mass shell, and summing over the spins of the 
final particles, the differential decay distribution  of
${B^0}\to K^{*0}(\to K^- \pi ^+)l^+l^-$  can be written as\footnote{For 
a $K\pi$ pair with an invariant mass $s_{K\pi}\neq m_\kstar^2$, the decay is 
parametrized by five kinematic variables.}
\bea\label{diff:four-fold}
d^4\G= \frac{9}{32 \pi} I(s, \theta_l, \theta_{K^*}, \phi) ds\,
d\cos\theta_l\,  d\cos\theta_{K^*}\, d\phi,
\eea
with the physical region of phase space 
\bea\label{int:region}
4 m_l^2\leqslant s\leqslant (m_B-m_{K^*})^2,\quad
-1\leqslant\cos\theta_l\leqslant 1,\quad 
-1\leqslant\cos\theta_{K^*}\leqslant 1, \quad
0\leqslant\phi\leqslant 2\pi,
\eea 
and
\bea\label{funcs:i}
I &=& I_1 + I_2\cos 2\theta_l + I_3 \sin^2\theta_l\cos 2\phi + I_4 \sin 2\theta_l \cos\phi + I_5 \sin\theta_l\cos\phi 
+ I_6 \cos\theta _l \nnu\\
&+&  I_7 \sin\theta_l\sin\phi + I_8 \sin 2\theta_l \sin\phi + I_9 \sin^2\theta_l\sin 2\phi.
\eea
The three angles $\theta_l, \theta_{K^*}, \phi$, which uniquely describe the decay 
${B^0}\to K^{*0}(\to K^- \pi ^+)l^+l^-$, are illustrated 
in \fig{fig:kinematics}. %
%
%
%
%
Note that $\phi$ is the angle between the normals to the planes defined by $K^-\pi^+$ and $l^+l^-$ in the rest frame of the $B$ meson; that is, defining the unit vectors
\be
{\bf e}_l=\frac{{\bf p}_{l^-}\times{\bf p}_{l^+}}{|{\bf p}_{l^-}\times{\bf p}_{l^+}|}, 
\quad {\bf e}_K=\frac{{\bf p}_{K^-}\times{\bf p}_{\pi^+}}{|{\bf p}_{K^-}\times{\bf p}_{\pi^+}|}, 
\quad {\bf e}_z=\frac{{\bf p}_{K^-}+{\bf p}_{\pi^+}}{|{\bf p}_{K^-}+{\bf p}_{\pi^+}|},
\ee 
where ${\bf p}_i$ denote three-momentum vectors in the 
$B$ rest frame, we have
\be
\sin\phi= ({\bf e}_l\times {\bf e}_K)\cdot {\bf e}_z , 
\quad \cos\phi= {\bf e}_K\cdot {\bf e}_l.
\ee

The functions $I_{1-9}$ in \eq{funcs:i} can be written in terms of the 
transversity amplitudes $\al$, $\ap$, $\app$, $A_t$. The last of these
corresponds to the scalar component of the virtual $K^*$, which is
 negligible if the lepton mass is small.
For $m_l\neq 0$, we find
\begin{subequations}\label{Isubis}
\bea
I_1 &=&  \bigg\{\frac{3}{4}[|\appl|^2 + |\apl|^2 + (L\to R)]\bigg(1-\frac{4m_l^2}{3 s}\bigg)
+ \frac{4 m_l^2}{s}\Re(\appl^{}\appr^* + \apl^{}\apr^*) \bigg\}
\sin^2\theta_{K^*}\nnu\\
&+& \bigg\{(|\all|^2 +|\alr|^2 ) + \frac{4m_l^2}{s}[|A_t|^2 + 2\Re(\all^{}\alr^*)]\bigg\}
\cos^2\theta_{K^*},
\eea
\bea
I_2 &=& \bigg(1 - \frac{4 m_l^2}{s}\bigg)\bigg[ \frac{1}{4}( |\appl|^2+ |\apl|^2)\sin^2\theta_{K^*}-
|\all|^2\cos^2\theta_{K^*}  + (L\to R)\bigg],
\eea
\bea
I_3 = \frac{1}{2}\bigg(1 - \frac{4 m_l^2}{s}\bigg)\bigg[(|\appl|^2 - |\apl|^2 )\sin^2\theta_{K^*} + (L\to R)\bigg],
\eea
\bea
I_4 =  \frac{1}{\sqrt{2}}\bigg(1 - \frac{4 m_l^2}{s}\bigg)\bigg[\Re
(\all^{}\apl^*) \sin 2\theta_{K^*}
+ (L\to R)\bigg],
\eea
\bea
I_5 = \sqrt{2}\bigg(1 - \frac{4 m_l^2}{s}\bigg)^{1/2}\bigg[\Re(\all^{}\appl^*) \sin2\theta_{K^*} - (L\to R)\bigg],
\eea
\bea
I_6 = 2\bigg(1 - \frac{4 m_l^2}{s}\bigg)^{1/2} \bigg[\Re (\apl^{}\appl^*) \sin^2\theta_{K^*} - (L\to R)\bigg],
\eea
\bea
I_7 =  \sqrt{2} \bigg(1 - \frac{4 m_l^2}{s}\bigg)^{1/2}\bigg[\Im (\all^{}\apl^*) \sin2\theta_{K^*} - (L\to R)\bigg],
\eea
\bea
I_8 =  \frac{1}{\sqrt{2}}\bigg(1 - \frac{4 m_l^2}{s}\bigg)\bigg[\Im
(\all^{}\appl^*) \sin2\theta_{K^*}
+ (L\to R)\bigg],
\eea
\bea
I_9 = \bigg(1 - \frac{4 m_l^2}{s}\bigg)\bigg[\Im (\apl^{*}\appl) \sin^2\theta_{K^*} + (L\to R)\bigg].
\eea
\end{subequations}%
The expression for the differential decay rate in \eq{diff:four-fold}, together with the formulae in Eqs.~(\ref{Isubis}), 
agrees with the result derived in  \rf{4body:mass} for the decay $B\to D^*
(\to D\pi)l^+l^- $  and  $m_l\neq 0$,\footnote{We take into account some 
obvious misprints in Eq. (37) of \rf{4body:mass}.}
and with \rf{FK:etal} in the case of
massless leptons (see also \rfs{dmitri, CS:etal}).
The differential decay rate in terms of the helicity amplitudes
$H_{\pm 1}, H_0, H_t$ 
can be obtained from Eqs.~(\ref{Isubis}) by means of the relations 
in \eq{hel:trans}.
%
%
%
%

\end{document}